\documentclass[12pt,preprint]{aastex}

\def\gtorder{\mathrel{\raise.3ex\hbox{$>$}\mkern-14mu
                \lower0.6ex\hbox{$\sim$}}}
\def\ltorder{\mathrel{\raise.3ex\hbox{$<$}\mkern-14mu
                \lower0.6ex\hbox{$\sim$}}}
\newcommand{\hii}{H~{\small II} }
\newcommand{\bra}{Br$\alpha$}
\newcommand{\brb}{Br$\beta$}
\newcommand{\bracket}{Br$\alpha$/Br$\beta$}
\newcommand{\neii}{[Ne~{\small II}]12.8$\mu$m}
\newcommand{\neiii}{[Ne~{\small III}]15.6$\mu$m}
\newcommand{\neiiii}{[Ne~{\small III}]36.0$\mu$m}
\newcommand{\neon}{[Ne~{\small III}]15.6$\mu$m/[Ne~{\small II}]12.8$\mu$m}
\newcommand{\arii}{[Ar~{\small II}]6.99$\mu$m}
\newcommand{\ariii}{[Ar~{\small III}]8.99$\mu$m}
\newcommand{\argon}{[Ar~{\small III}]8.99$\mu$m/[Ar~{\small II}]6.99$\mu$m}
\newcommand{\siii}{[S~{\small III}]18.7$\mu$m}
\newcommand{\siiii}{[S~{\small III}]33.5$\mu$m}
\newcommand{\sulfur}{[S~{\small III}]18.7$\mu$m/[S~{\small III}]33.5$\mu$m}
\newcommand{\sulfura}{[S~{\small IV}]10.5$\mu$m/[S~{\small III}]33.5$\mu$m}
\newcommand{\siv}{[S~{\small IV}]10.5$\mu$m}

\begin{document}

\title{The Excitation and Metallicity of Galactic H~{\normalsize\bf II}
Regions from ISO-SWS Observations of Mid-Infrared Fine-Structure Lines
  \footnote{Based on observations with ISO, an ESA project with
    instruments funded by ESA Member States (especially the PI
   countries: France, Germany, the Netherlands, and the United Kingdom) with
the participation of ISAS and NASA.}}
\author{Uriel Giveon and Amiel Sternberg}
\vspace{0.5cm}
\affil{School of Physics and Astronomy and the Wise Observatory,
        The Beverly and Raymond Sackler Faculty of Exact Sciences,
        Tel Aviv University, Tel Aviv 69978, Israel; giveon@wise.tau.ac.il,
	amiel@wise.tau.ac.il}
\vspace{0.5cm}
\author{Dieter Lutz and Helmut Feuchtgruber}
\vspace{0.5cm}
\affil{Max-Planck-Institut f\"{u}r Extraterrestrische Physik,
	Giessenbachstra\ss\/e, 85748 Garching, Germany; lutz@mpe.mpg.de,
	fgb@mpe.mpg.de}
\vspace{0.5cm}
\and
\vspace{0.5cm}
\author{A.W.A. Pauldrach}
\vspace{0.5cm}
\affil{Institute for Astronomy and Astrophysics of the Munich
	University, Scheinerstra\ss\/e 1, D-81679 Munich, Germany;
       uh10107@usm.uni-muenchen.de}
\vspace{1cm}

\begin{abstract}
We present mid-infrared ISO-SWS observations of the fine-structure emissions
lines \neii, \neiii, \neiiii, \arii, \ariii, \siii, \siiii, \siv\, and the
recombination lines \bra\ and \brb, in a
sample of 112 Galactic \hii regions and 37 nearby extra-galactic \hii
regions in the LMC, SMC, and M33. We selected our sources from archival
ISO-SWS data as those showing prominent \neii\ or \neiii\ emissions. The
Galactic sources have a wide range in galactocentric distance (0 kpc $\ltorder
R_{gal}\ltorder~18$ kpc), which enables us to study excitation and metallicity
variations over large Galactic scales. We detect a steep rise
in the \neon, \argon, and \sulfura\ excitation ratios from
the inner Galaxy outwards, and a moderate decrease in metallicity,
from $\sim 2Z_{\odot}$ in the inner Galaxy to $\sim 1Z_{\odot}$ in
the outer disk. The extra-galactic sources in our sample
show low gas density, low metallicity and high excitation. 
We find a good correlation between \neon\ and \argon\ excitation ratios in our
sample. The observed correlation is well-reproduced by theoretical nebular
models which incorporate new generation wind-driven NLTE model stellar
atmospheres for the photoionizing stars.
In particular, the NLTE atmospheres can account for the
production of [Ne~{\small III}] emission in the \hii regions.
We have computed self-consistent nebular and stellar atmosphere
models for a range of metallicities (0.5 to 2 $Z_{\odot}$).
We conclude that the increase in nebular excitation with galactocentric
radius is due to an increase in stellar effective temperature
(as opposed to a hardening of the stellar SEDs due to the metallicity
gradient).
We estimate an integrated \neon\ ratio for the Galaxy of 0.8, which puts it
well inside the range of values for starburst galaxies. The
good fit between observations and our models support the conclusion of
Thornley et al. (2000), that the low \neon\ ratios observed in extra-galactic
sources are due to global aging effects.
\end{abstract}


\section{Introduction}
\label{introduction}

Mid-infrared (IR) fine-structure emission lines are useful probes of the
physical properties of \hii regions and their associated ionizing sources.
Unlike optical and UV emission lines, the fine-structure lines do not suffer as
much from dust extinction, and 
the low energies of the fine-structure levels ($h\nu/k\sim 10^{2-3}$K), make
the fine-structure lines relatively insensitive to
the nebular electron temperature $T_e$ ($\sim 10^4$K in \hii regions).
Fine-structure line emissions from photoionized gas
have been observed in a wide range of 
astrophysical sources (Genzel et al. 1998; Cox et al. 1999; Thornley et
al. 2000). Their measured intensities have been shown to be strong,
comparable to optical hydrogen recombination lines. 
The relative strengths of the fine-structure emission
lines may be used to
probe the ionizing spectral energy distributions (SEDs)
because different ionization stages are
formed by significantly different photon energy ranges in the ionizing
spectrum. For example, the \neon\ line ratio is
sensitive to photons emitted at $\gtrsim 3$ Rydberg
region, while \argon\  probes the region $\gtrsim 2$
Rydberg. Measuring these (and other) ratios can constrain the SEDs
of the hot stars in \hii regions (Thornley et al. 2000;
F\"{o}rster Schreiber et al. 2001a) or of the non-stellar sources
in active galactic nuclei (Alexander et al. 1999; 2000).

The Short Wavelength Spectrometer (SWS) on board 
{\it Infrared Space Observatory} (ISO) made spectroscopic studies
of Galactic and extragalactic sources possible across
the entire mid-infrared (2.5--45$\mu$m) spectral range
(de Graauw et al. 1996; Kessler et al. 1996; Genzel \& Cesarsky 2000).
In this paper we have compiled SWS observations of fine-structure
emission lines from Galactic and nearby extra-galactic \hii regions
from the ISO data archive. The sources were selected as those
in which the \neii\ or \neiii\ lines are present,
and eliminating target types other 
than \hii regions. Since these objects are typically prominent sources
of the diagnostically useful \arii, \ariii, \siii, \siiii, and \siv\
fine-structure lines and the \bra\ and \brb\ recombination lines, we extracted
these lines as well for sources in which they were observed. 

In this paper we have several goals. First, we wish to re-investigate
the large scale variations in elemental abundances and nebular excitation
across the Galactic disk. The Galactic sources in
our sample have a wide enough range in galactocentric distance (0~kpc~$\ltorder
R_{gal}\ltorder 18$~kpc) to enable us to study these variations.
Such studies using ISO data have been presented by Cox et
al. (1999) for a sample of 46 Galactic \hii regions over a
range of 0 kpc to 25 kpc in galactocentric radius, and showed
that the excitation ratios \neon\ and \argon\ increase as a function
of the galactocentric radius. They also showed that the argon and neon
abundances, based on observations of \neii\ and \arii\, decrease
by a factor of 2 from the inner Galaxy outwards
(see also Roelfsema et al. 1999 and
Mart\'{\i}n-Hern\'{a}ndez 2000). These results are consistent
with the trends found in earlier studies of Galactic
\hii regions using ground-based and
air-borne observations (Simpson, Colgan, \& Rubin 1995; Afflerbach, Churchwell,
\& Werner 1997).
Here we present results for a larger source sample.
Understanding the excitation gradient 
within our galaxy may shed further light on excitation
variations observed in external galaxies
(e.g. Shields \& Tinsley 1976 in M101; Henry \& Howard 1995 in M33, M81
and M101; see also Shields 1990).

Second, we use the
observations as templates for constraining the results of new generation wind-
driven (NLTE)
stellar atmosphere computations (Pauldrach, Hoffmann, \& Lennon 2001).  
For example, a long-standing problem in Galactic \hii regions studies
has been the failure of photoionization simulations
employing old-generation static (LTE) atmospheres (Kurucz 1992)
to reproduce the observed (high-excitation) [Ne~{\small III}] line emission
(Baldwin et al. 1991; Simpson et al. 1995). We re-investigate
(and resolve) this issue here using NLTE atmospheres. 
In our analysis we construct an \argon\ {\it vs.\/} \neon\ diagnostic diagram
as a measure of the nebular excitation. 
We fit the observed correlation between these two line ratios by constructing
nebular simulations in which the photoionizing stellar SEDs
are modeled using NLTE atmosphere computations. 
We compare with results obtained using the older LTE Kurucz atmospheres.
We also explore the effects of metallicity variations 
by self-consistently varying the nebular and the stellar atmosphere abundances.

This paper is a part of a larger framework, whose aims are to study
theoretical models of starburst regions, as they are probed by IR
spectroscopy (Kunze et al. 1996; Lutz et al. 1996; Rigopoulou
et al. 1996; Sturm et al. 1997; Genzel et al. 1998; Thornley et al. 2000,
F\"{o}rester Schreiber et al. 2001a,b).
For example, Thornley et al. (2000) measured the \neon\ excitation
ratio in extra-galactic starburst sources, and found it to be
{\it low} compared to starburst models which incorporate the
new-generation atmospheres. They concluded that this is evidence
for global aging of the star-forming regions.
Fine-structure emission lines have also been used to try and
determine the primary energy source in the ultra-luminous
infrared galaxies (ULIRGs; Lutz et al. 1996; Genzel et al. 1998;
Rigopoulou et al. 1999).
Understanding the properties of the SEDs of hot stars, as
probed by Galactic \hii regions, is crucial to the interpretation of
the extra-galactic observations.

\section{Observations}
\label{observations}

\subsection{Data Reduction and Source Selection}
\label{sample}

The ISO data archive contains several thousands of observations that
were obtained with the 2.4--45$\mu$m grating spectrometer SWS. We have 
identified potential \hii regions in all observations taken in the
full spectrum mode (SWS01) and the line spectroscopy mode (SWS02) during
the ISO mission. An automatic procedure starting from the off-line processing 
version 7 standard processed data (SPD) processed all these observations to
the auto analysis result (AAR) level. From every AAR, a spectral range of
$\pm$30 resolution elements (r=2000) around the \neii\ and 
\neiii\ lines, was cut out for further analysis.
This processing was automated by a script which flat-fielded the 
individual detectors to their mean value, removed 3$\sigma$ outliers and 
rebinned each line to a single-valued spectrum.

For the SWS01 data, a simple automatic removal of the SWS band-3
instrumental fringes, was applied in this procedure
when the continuum for the \neii\ and \linebreak
\neiii\ lines exceeded 100 Jy. 
We then automatically fitted gaussians to the rebinned spectral segments and
measured line fluxes. A source was considered a candidate for in-depth 
processing if at least one of the two lines was apparently present 
at a level above $10^{-20}$ W\,cm$^{-2}$.

For the SWS02 data, the automatically produced single valued spectral 
segments for each line were stored and later inspected visually
for the presence of lines, which finally gave a list of observation number 
and target name for every clear detection of either \neii\, or \neiii\, or 
both. Visual inspection was preferred here, since the defringing step 
in AOT02 observations is less robust, and a lot of faulty detections 
would occur otherwise.

Our selection procedure initially gave also objects of similar spectral
characteristics (nearby galaxies, planetary nebulae etc.) which were 
subsequently identified and excluded from further processing. All observations
from the list of candidate \hii regions were then reprocessed
and the line fluxes were measured.
In Table~\ref{linelist} we list the the specific emission lines included in our
source sample, together with their ionization potentials, line wavelengths,
and critical densities for electron impact collisions.

For the SWS01 data, this needed special care since a large fraction of the
observations was taken in the faster speed options of the SWS01 mode.
For those observations, the grating moved during integrations causing curved
integration ramps when passing a strong emission line. The automatic 
deglitching of the SWS data occasionally confused such curved
ramps with jumps in the ramps that correspond to glitches caused by ionizing
particle hits, and thus might have affected emission line fluxes (Section 4.5
of ISO Handbook Volume 6, The Short Wavelength Spectrometer, SAI-2000-008/Dc, 
Version 1.1., http://www.iso.vilspa.esa.es/manuals/HANDBOOK/VI/sws\_hb/). 
All SWS01 data except for the slowest speed 4 were hence reprocessed twice from
ERD level, once with standard OLP Version 8 calibration files and once
with one calibration file changed to switch off the glitch recognition.
In both cases, the data were processed to the AAR level, cleaned from
outliers by iterative clipping, flat-fielded, and rebinned to a
single-valued spectrum. The two versions were then plotted and
inspected to determine whether the deglitching might have affected the
real line in the standard processing, or whether a glitch might have
been present in those data with glitch recognition
switched off. This classification was in most cases possible on the basis 
of feature shape and widths. The line flux or limit was then measured in that
version not affected by problems. In a few cases with both versions affected,
generous upper limits were adopted. SWS01 speed 4 and SWS02 data were
processed by default before measuring line fluxes by direct integration of the
continuum subtracted line, or determining upper limits corresponding to a
line, of 3$\sigma$ peak height and the proper width. For SWS01 observations
this was done for all lines in Table~\ref{linelist}, for SWS02 only
for those present in a particular observation. The uncertainty in the line
fluxes is estimated to be $\sim 30\%$.

The final identification of individual
objects -- whether it is an \hii region or a different kind of
source (e.g. nearby galaxy, planetary nebula etc.) that was missed
by the first filtering, or whether there were multiple observations of
the same object -- is based on a literature survey, mainly
using the SIMBAD astronomical database\footnote{http://simbad.u-strasbg.fr/}.

A total of 149 SWS spectra were selected for the final sample. This
includes almost all of the \hii regions observed by ISO-SWS - 112
Galactic \hii regions, spanning galactocentric
radii of 0 to $\sim$18 kpc, 14 \hii regions in the Large Magellanic Cloud
(LMC), 4 in the Small Magellanic Cloud (SMC), and 19 in the nearby spiral
galaxy M33.
This sample is certainly heterogeneous since it is a collection of targets
observed for different proposals and scientific reasons. These include
detailed studies of well known individual \hii regions, more 
systematic surveys of compact \hii regions originally selected on the basis 
of IRAS-LRS spectra and on the basis of IRAS sources located outside the 
solar circle, observations of the brightest \hii regions in the 
LMC, SMC, and M33, and \hii region observations targeted at specific 
issues of ISM physics rather than a per se characterization of the 
\hii region observed. The selection criteria adopted by these
proposals necessarily differed, and incomplete execution of planned 
observations during the ISO mission induces another (though uncritical) 
selection effect. Some observations represent integrated fluxes of rather
compact well-defined \hii regions, others different parts of more complex 
regions, including pointings in the outskirts of resolved nebulae.
However, the sample spans \hii regions with a wide enough range of positions
and excitation conditions to study both the stellar SEDs and 
Galactic metallicity and nebular excitation gradients.

Metallicity gradient analyses in our Galaxy should be little affected by 
possible selection biases in our sample, as long as there is efficient 
mixing at a given galactocentric radius. The situation is less clear 
for excitation analyses where competing selection biases are possible. 
Selection on compactness, for example, might prefer younger (harder) 
but also less luminous (softer) objects, with additional effects due to
dust possible. Similarly, optical rather than infrared/radio selection 
might prefer more evolved (softer) but also more luminous (harder) 
and lower metallicity (harder) objects, all perhaps more likely to be 
optically visible. Since a large fraction of the targets in the ISO 
database is originally IRAS-selected, such effects are likely not 
governing excitation analyses as a function of galactocentric radius in 
our sample.

\subsubsection{Calculation of Galactocentric Radius}
\label{distances}

We obtained radial velocity measurements for most of the Galactic
sources from the literature (see references in Tables~2
and \ref{data_ref}). These velocities were determined using different methods
and are based on various emission features (e.g., radio recombination lines;
sub-mm molecular lines of CO, CS, and CI; maser emission). In some cases (DR21,
GAL 007.47+00.06), the given velocity value is the mean between different
values obtained from the same method (i.e., several recombination lines).

We use the Rohlfs \& Kreitschmann (1987) Galactic rotation curve
to calculate the galactocentric distance
of each source. Their rotation curve spans a range of 0.1 to 19.4 kpc in
galactocentric radius. For the Sun we adopt a galactocentric distance
$R_{\odot}=8.5$ kpc and rotational velocity $\theta_{\odot}=220$ km s$^{-1}$
(Honma \& Sofue 1997). We have transformed the rotation curve of Rohlfs
\& Kreitschmann ($R_{\odot}=7.9$ kpc and $\theta_{\odot}=184$ km s$^{-1}$)
using the more updated solar values of Honma \& Sofue.
Most of the Galactic sources lie very near the
Galactic plane (typically, $b<5^o$), so we assume that they lie in the
Galactic plane ($b=0$). (An exception is the Orion nebula ($b=-19.4^o$),
for which the following procedure was not applied.)
For each source having a
radial velocity $v_{los}$ and a Galactic longitude $l$, we compute
the quantity $W(R_{gal})$, where $v_{los}=W(R_{gal})\sin{l}$, and
\begin{equation}
W(R_{gal})=\left[\theta(R_{gal})\left({{R_{\odot}}\over {R_{gal}}}\right)-\theta_{\odot}\right].
\label{rot_curve_eq}
\end{equation}
Here, $R_{gal}$ is the galactocentric radius and $\theta(R_{gal})$ is the
rotational velocity at $R_{gal}$.
Using a linear interpolation of the $W(R_{gal})$ curve from Honma \& Sofue
we find the values of $R_{gal}$ corresponding to the $W(R_{gal})$ of
each source.

The kinematic distance ($R_{kin}$; distance from the Sun) is calculated by
inverting the formula
\begin{equation}
R_{gal}^2=R_{\odot}^2-2R_{kin}R_{\odot}\cos{l}+R_{kin}^2.
\label{r_gal_eq}
\end{equation}
For sources outside the solar circle, this gives a single solution, whereas
inside the solar circle, one obtains two solutions. The choice of distance in
this case is based on different methods discussed in the references. A distance
ambiguity remains for only three sources designated with asterisks in Table 2
(column 8).

\subsubsection{The Sources Data}
\label{lit_data}

Table 2 lists the data collected from the literature on
the Galactic sources.
The columns in this table give the following information:
\begin{itemize}
\item[]{\it Column (1)\/}---Source identification number in the sample. These
  numbers also appear as source identifications in the various figures.
\item[]{\it Column (2)\/}---Source name.
\item[]{\it Column (3-4)\/}---Position (J2000) of the ISO aperture in Right
Ascension and declination.
\item[]{\it Column (5)\/}---Kinematic distance (distance from the Sun) in kpc.
\item[]{\it Column (6)\/}---Galactocentric radius (see \S~\ref{distances}) in
kpc.
\item[]{\it Column (7)\/}---Source radial velocity in km s$^{-1}$.
\item[]{\it Column (8)\/}---Acronym for the distance/radial velocity
  reference. The literature references are given in Table~\ref{data_ref}
\item[]{\it Column (9-18)\/}---ISO-SWS fluxes for the lines listed in
  Table~\ref{linelist}. The fluxes are in units of $10^{-19}$ W cm$^{-2}$. The
error in the flux measurement is $\sim 30\%$.
\end{itemize}

In several cases, where observations of different parts of 
spatially resolved and complex regions were available,  we
summed the multiply observed intensities across the regions to
a single value for each line.
These observations are listed in Table~\ref{multiple_obs}.
Table 5 lists the data for sources in the LMC and SMC.
In Table 6 we list the data for sources in M33. In this table
we also list the angular distance of each source from the center of the M33.

A few studies of individual sources in our sample have been
presented in the literature. For these sources our fluxes
are in good agreement with the previously published fluxes.
For example, in Orion IRC2 van Dishoeck et al. (1998) give fluxes for all of
our lines  (except for \brb) which are similar to the fluxes of our source
\#13.
Similarly, van den Ancker, Tielens, \& Wesselius (2000) published fluxes for
the \neii, \neiii, \siii, \siiii, and \siv\ fine-structure lines of S106 IR,
which are comparable to those of our S106 IRS4 (\#99).
Rosenthal, Bertoldi, \& Drapatz (2000) published fine-structure fluxes for
all of the lines we use in this paper for the Orion $H_2$ emission peak 
nebula (our sources \#11 and \#12).
Rodr\'{i}guez-Fern\'{a}ndez, Mart\'{i}n-Pintado, \& de Vicente (2001) give
fluxes of several sources in the Galactic center from which six sources are
included in our sample (\#62-\#67). They measured \neii, \neiii, \siii, and
\siiii.

\section{Results}
\label{results}

\subsection{Excitation}
\label{excitation}

\begin{figure}[p]
\plotone{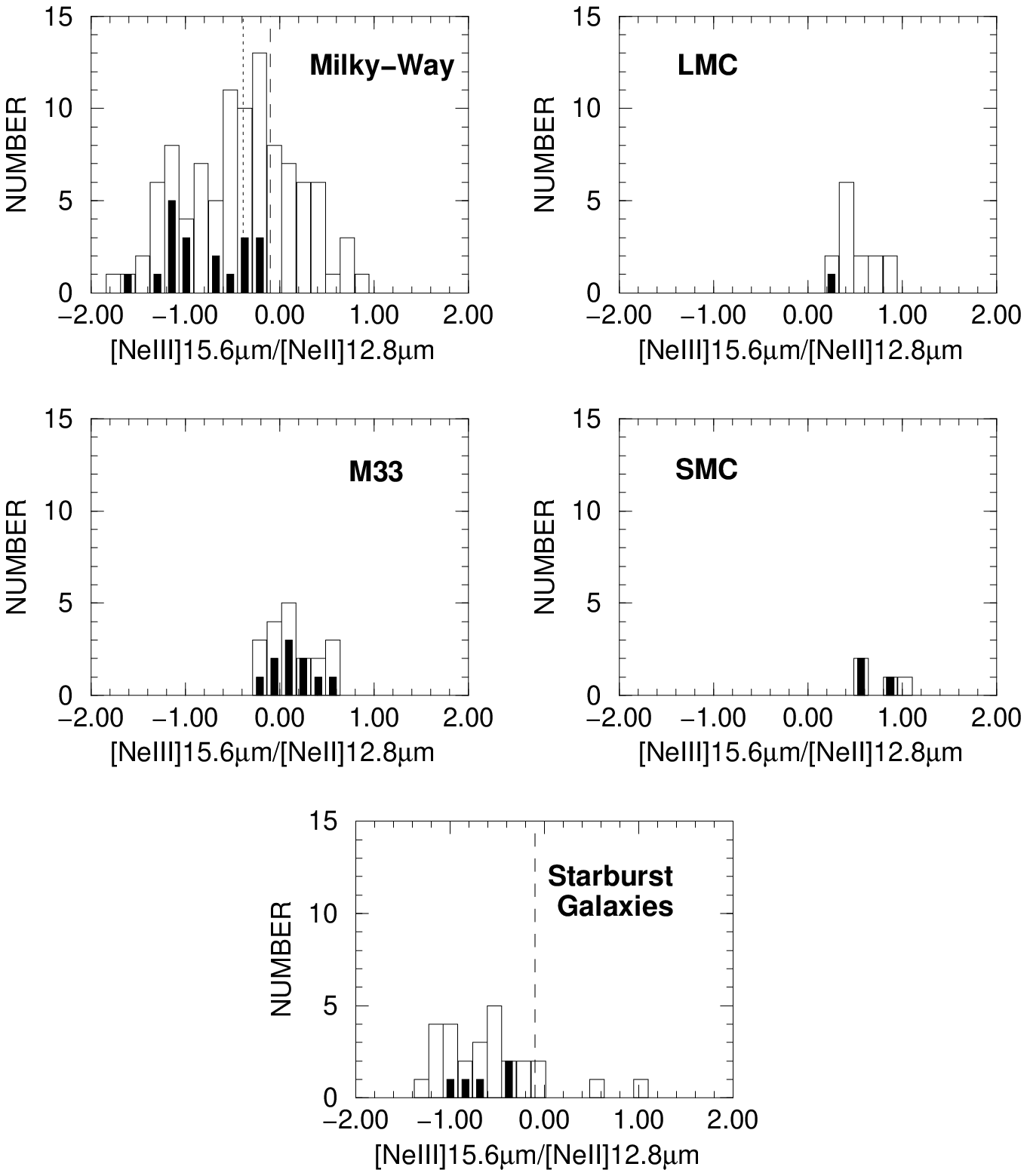}
\caption[Histograms of the observed logarithmic \neon\ ratio
(not corrected for extinction) for sources in the our sample -- the Galaxy,
the LMC, the SMC and M33, and also of extra-galactic sources from Thornley et
al (2000). Filled bars designate sources whose ratios are uncertain. For the
Milky-Way and the starburst sample, these are all upper limits, for the LMC and
SMC - lower limits, whereas in M33 both cases are mixed together.
Our estimation of the integrated \neon\ ratio for the
Galaxy is over-plotted as vertical long-dashed lines both in the Galactic
and extra-Galactic histograms. The median \neon\ ratio for the Galaxy
is shown as the vertical short-dashed line in the histogram of the
Galaxy. The Galaxy appears to have higher
excitation ratio than most starburst galaxies.]{}
\label{ne_hist_all}
\end{figure}

We use the line ratios \neon, \argon, and \sulfura\ to quantify the
source excitation. In Figure~\ref{ne_hist_all} we show histograms of
the observed \neon\ ratio (not corrected for extinction) for the groups
of sources in our sample -- the Milky Way, 
the LMC, the SMC, and M33, and for comparison, the
sample of extra-galactic sources, as measured by Thornley et al. (2000).
Filled bars designate sources whose ratios are uncertain. For the Milky-Way and
the starburst sample these are all upper limits, for the LMC and SMC they are
lower limits, whereas in M33 both cases are mixed together.
It appears that the excitation distribution of Galactic \hii
regions is quite different from the distribution in the LMC, SMC and
M33, with smaller excitation in the Milky Way on average.

We can use our observations of Galactic \hii regions to estimate an integrated
value of the \neon\ ratio for our Galaxy which can be compared to the starburst
galaxies. This is necessarily approximate since the sample drawn from the
ISO archive is not an unbiased representation of the nebular luminosity
of the Galaxy. The integrated ratio (not corrected for extinction) equals 0.8,
with a de-reddened value (see \S 3.2) of 0.7. 
The corresponding median values are 0.41 and 0.36.
These, admittedly approximate values (shown as vertical lines
over-plotted on the Galactic and extra-galactic histograms in
Figure~\ref{ne_hist_all}) put our Galaxy well inside the 
range of \neon\ values for much more active starburst galaxies observed 
by Thornley et al. (2000), and even above its median. 

\begin{figure}[p]
\plotone{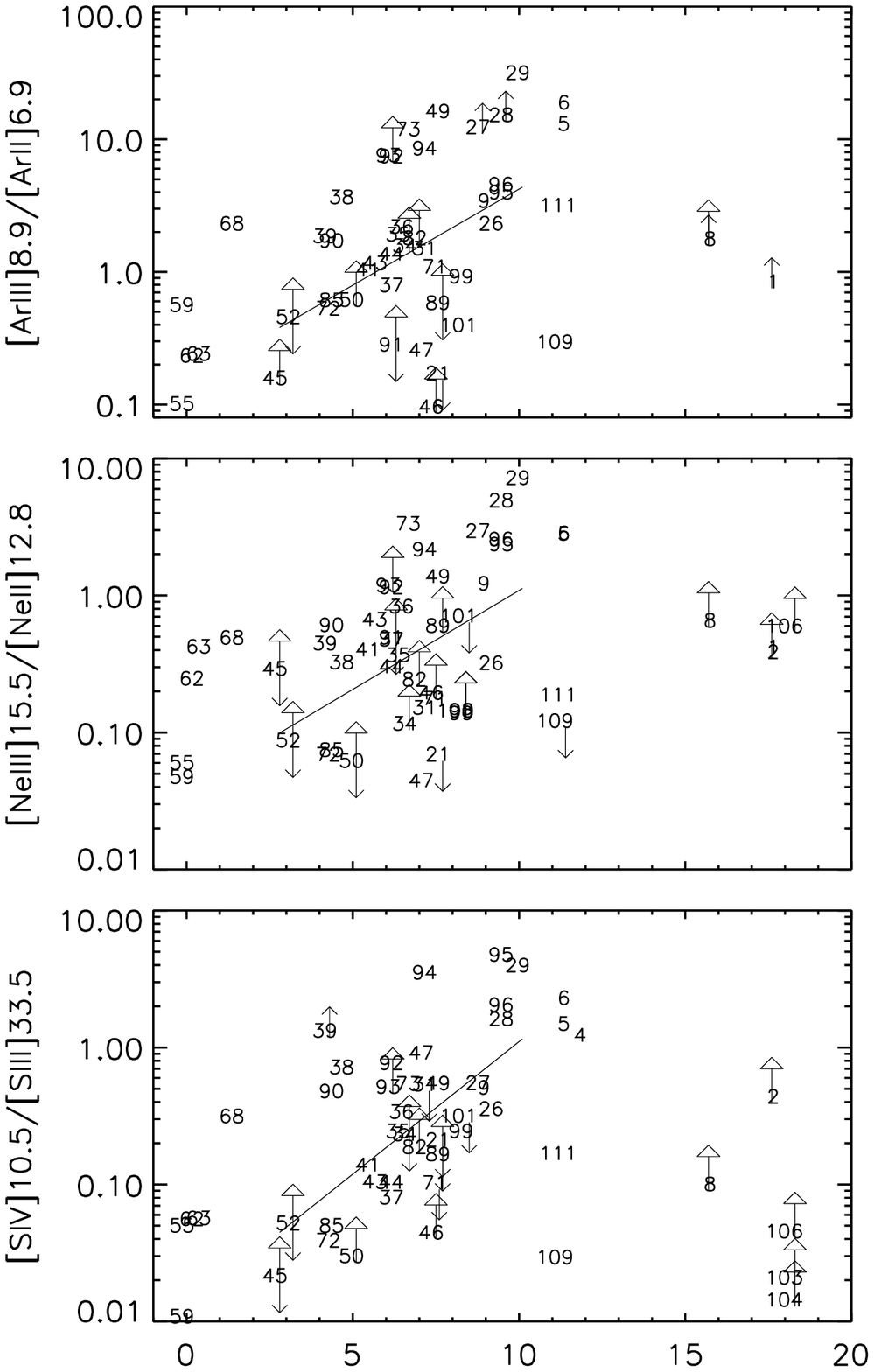}
\vspace{25cm}
\caption[Excitation ratios {\it vs.} galactocentric radius for
  Galactic sources. The plot for \argon\ is in the upper panel. In the middle
panel we show the \neon\ plot, and \sulfura\ is in the lower panel. A rise
in the excitation towards the outer Galaxy is clear. Solid lines are fits to
the gradients (for $2\leq R_{gal}\leq 11$ kpc), whose values are given in the
text. We designate limits on the
given quantities by small arrows. The larger arrows indicate lower limits
resulting from uncertainties in the extinction correction.]{}
\label{excite_rgal}
\end{figure}

In Figure~\ref{excite_rgal} we plot the three excitation ratios
mentioned above {\it vs.\/} galactocentric radius for the Galactic sources.
The data points are
represented by the source identification numbers, as listed in Table 2.
The displayed ratios have been corrected for extinction (\S~\ref{extinction}).
In this figure and the figures that follow we designate limits on the
given quantities by small arrows. Large arrows designate lower
limits on the extinction correction.

Figure~\ref{excite_rgal} shows a clear rise in the excitation towards
the outer Galaxy. We have fitted power-law curves to the data in the range
$2\leq R_{gal}\leq 11$ kpc. The fitted gradients are
\begin{equation}
\log{{{{\rm [Ne\ III]}15.6\mu m}\over{{\rm [Ne\ II]}12.8\mu
      m}}}=0.14(\pm 0.05)R_{gal}-1.4(\pm 0.3),
\label{neon_grad}
\end{equation}
\begin{equation}
\log{{{{\rm [Ar\ III]}8.99\mu m}\over{{\rm [Ar\ II]}6.99\mu m}}}=0.15(\pm
0.05)R_{gal}-0.8(\pm 0.4),
\label{argon_grad}
\end{equation}
and
\begin{equation}
\log{{{{\rm [S\ IV]}10.5\mu m}\over{{\rm [S\ III]}33.5\mu m}}}=0.19(\pm
0.04)R_{gal}-1.9(\pm 0.3).
\label{sulfur_grad}
\end{equation}
The observed excitation gradients are similar to those found in earlier
studies of Galactic \hii regions (Simpson et al. 1995; Afflerbach et
al. 1997, Cox et al. 1999).
We note that the \neon\ and \argon\ gradients are
quite similar. This is an important clue as to the origin of the
excitation gradients as we discuss in \S~\ref{analysis}.

\begin{figure}[p]
\plotone{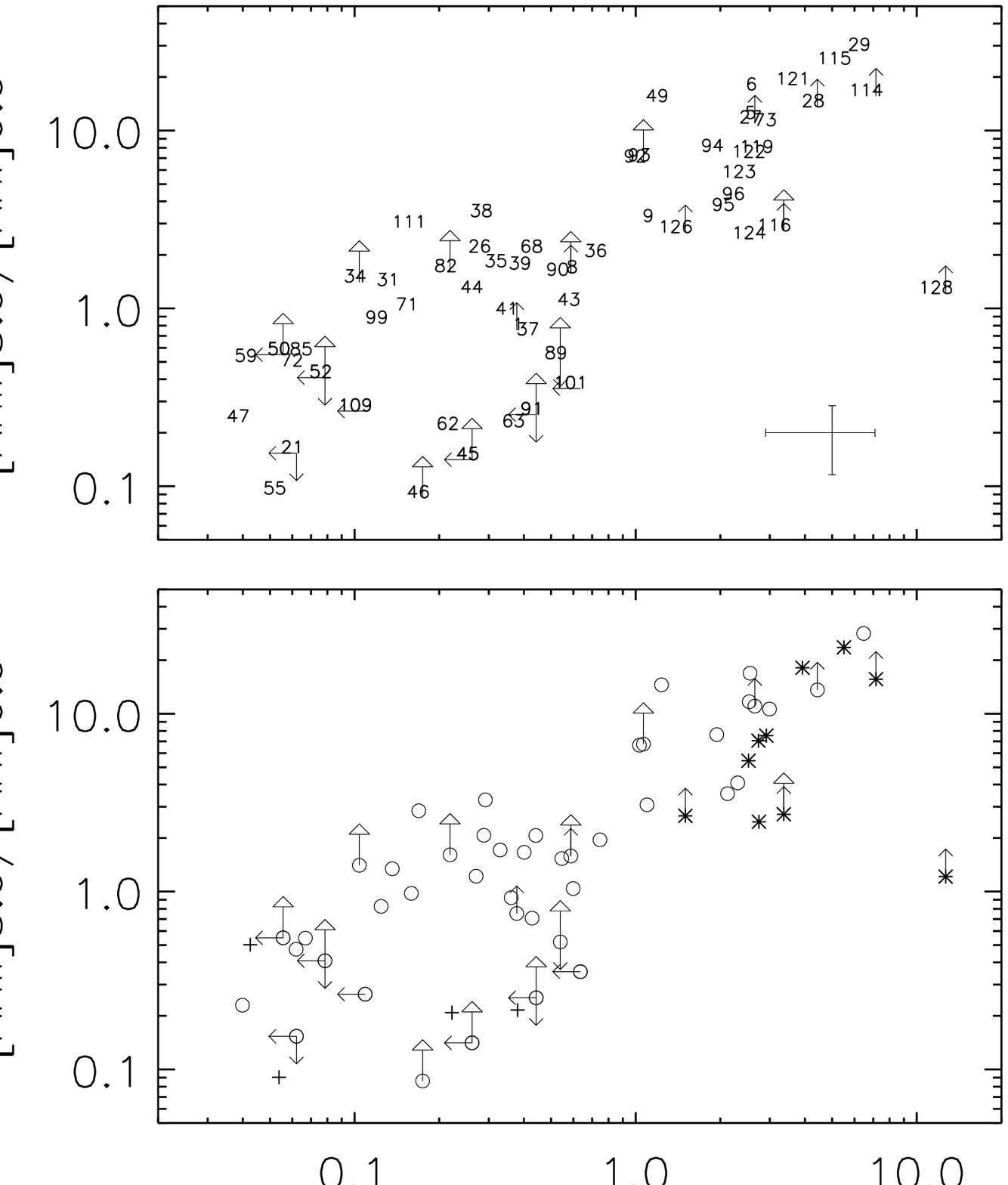}
\vspace{25cm}
\caption[Top panel - \argon\ {\it vs.} \neon\ for the entire
sample. The data points are represented by the source identifications
in Tables 2, 5 and 6. A representative error bar
for the data is shown in the lower right corner.
 Bottom panel - same as above, with data points marked with symbols
  according to their Galactocentric radius. Pluses designate the
  inner kpc in the Galaxy ($R_{gal}<1$ kpc), circles designate
  Galactic disk sources ($R_{gal}>1$ kpc), and asterisks designate
  extra-galactic sources.]{}
\label{ar_ne}
\end{figure}

The ionization potentials of Ar$^+$, S$^{++}$, and Ne$^+$ are equal to 27.63 eV
($\sim 2$ Rydberg), 34.79, and 40.96 eV ($\sim 3$ Rydberg), respectively (see
Figure~\ref{atmospheres}). Thus, the ratios \linebreak
\argon, \sulfura, and \neon\ 
constrain the shapes of the SEDs of the ionizing spectra over a range
of several Rydbergs. One possible representation for the data, which
may give clues regarding the properties of the SEDs, is a diagnostic
diagram combining two of these empirical ratios.
In Figure~\ref{ar_ne} we plot the \argon\ {\it vs.\/} \neon\ ratios
for our sample.
A good correlation is apparent, showing that nebular excitation
is probed by both line ratios, and manifests the inter-relation
between the two spectral regions, $2<\varepsilon <3$ Rydberg and
$\varepsilon>3$ Rydberg. A representative error bar 
for the data is shown in the lower right corner of the plot.

In order to connect this correlation in excitation space to position
within the Galaxy we display each source in the lower panel of
Figure~\ref{ar_ne} as a symbol according to its galactocentric radius. 
Pluses designate the inner kpc in
the Galaxy ($R_{gal}<1$ kpc), circles designate Galactic disk
sources ($R_{gal}>1$ kpc), and asterisks designate extra-galactic sources.
From this diagram it is clear that the sources in the central kpc of
the Galaxy, are all of low excitation. On the other hand, sources in
the LMC, are all of high excitation. Sources in the disk of the Galaxy
show moderate levels of excitation.

\subsection{Extinction and Gas Densities}
\label{extinction}

\begin{figure}[p]
\plotone{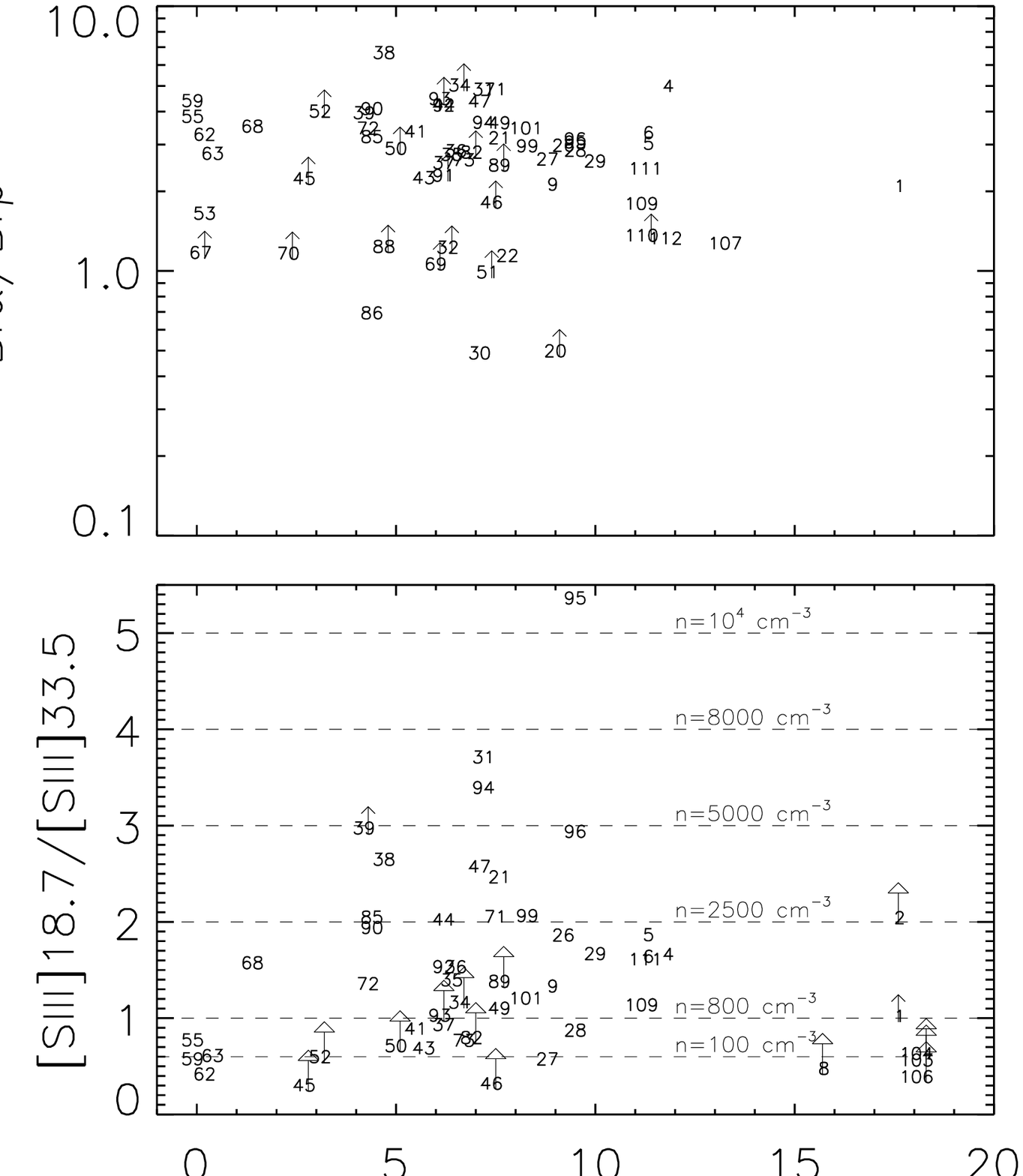}
\vspace{25cm}
\caption[The \bracket\ ratio (upper panel) and the \sulfur\
  ratio (lower panel) for Galactic sources in our sample that have
  measured fluxes for these lines, as a
  function of the galactocentric radius.]{}
\label{br_s3_rgal}
\end{figure}

We use the Brackett line ratio, \bracket\  (for those sources for
which both lines are measured) to de-redden the fine-structure line
intensities. We assume a foreground screen,
and adopt the Draine (1989) mid-IR extinction curve.
For case-B recombination the intrinsic Brackett
decrement ranges from 1.6 to 1.9 for gas temperatures
between $5\times 10^3$ and $2\times 10^4$ K,
and densities between $10^2$ and $10^4$ cm$^{-3}$. 
Figure~\ref{br_s3_rgal} shows the Brackett ratio for
the objects in our sample {\it vs.\/} galactocentric radius.

It is evident that there are objects with unphysical Brackett ratios much below
the optically thin ratio of $\sim 1.7$. These objects generally
correspond to sources with noisy low signal-to-noise spectra,
and we discard them in our subsequent analysis.
The extinction correction is most significant for the \ariii\ and \siv\ lines
which lie close to the mid-IR silicate absorption feature. 
For these lines correction factors as large as 10 to 25 are implied by
the Brackett decrements. The corrections for the other fine-structure
lines are small.
For some objects only an upper-limit exists for \brb. We adopt
the upper-limit values to compute a minimum extinction correction for such
objects. Objects with lower-limit extinction corrections are indicated by the
large arrows in the figures.

The \sulfur\ ratio was used as a gas density measure.
The solid angle of the SWS aperture used to observe the 33.5$\mu$m
line is 1.75 times larger than the one used for the 18.7$\mu$m
line. Depending on source structure, this may require aperture
corrections to the ratio up to about this factor. We have not
attempted to derive these corrections for individual sources but note
that observed ratios of $\approx$0.25 are possible for low
density extended sources.

In Figure~\ref{br_s3_rgal} we display the \sulfur\ {\it vs.\/} galactocentric
radius for the (extinction corrected) Galactic sources.
The measured densities for our sample lie in the range of $\ltorder 100$ to
$\sim 15,000$ cm$^{-3}$ with a mean value of 800 cm$^{-3}$.
Measurements of the S~{\small III} ratio are also available for some of
the LMC sources. These ratios lie in the range of $0.4-0.6$,
corresponding to densities $\ltorder 200$ cm$^{-3}$.

\subsection{Abundances}
\label{abundance_calc}

We have calculated the elemental abundances for neon and argon from the
relative strengths of the fine-structure and \bra\ emission lines.
In computing the fine-structure line emissivities we adopt the ``IRON project''
collision strengths (Saraph \& Tully 1994 for [Ne~{\small II}];
Butler \& Zeippen 1994 for [Ne~{\small III}];
Pelan \& Berrington 1995 for [Ar~{\small II}];
Galavis et al. 1995 for [Ar~{\small III}]). (These data are also
employed in the CLOUDY photoionization models we discuss in \S 4.)
The critical densities ($\gtrsim 10^5$ cm$^{-3}$)
of the relevant neon and argon transitions are much larger than the
estimated gas densities so that collisional de-excitations are negligible.
Because of the large ionization potential (4.7 Ryd) of Ne$^{++}$ we assume
that neon exists either as Ne$^+$ or Ne$^{++}$ in the \hii regions.
The ionization potential of Ar$^{++}$ is 3.0 Ryd, very similar to that
of Ne$^+$. However, Ar$^{+++}$ ions are rapidly removed by
charge transfer reactions with neutral hydrogen (Butler \& Dalgarno 1980).
In contrast, charge transfer neutralization is inefficient for Ne$^{++}$
(Butler, Heil, \& Dalgarno 1980). We therefore assume that
negligible [Ar~{\small IV}] is produced in the nebulae even
when [Ne~{\small III}]/[Ne~{\small II}] is large. This is verified by our
CLOUDY computations which incorporate the Butler et al. charge transfer
rate coefficients. Thus, we do not apply ionization ``correction
factors'' in deriving the neon and argon abundances from our
fine-structure line data.  

For the singly ionized species, Ar$^+$ and Ne$^+$, which have two
fine-structure levels in their ground state, the emissivity is given by
\begin{equation}
j_{\rm II}=n_e n q_{\rm II} h\nu_{\rm II} {{\rm X_{II}}\over{\rm H}},
\label{singly_ionized}
\end{equation}
where $n$ is the gas density,$n_e$ is the electron density, $q_{\rm II}$ is the
collisional excitation rate, and X$_{\rm II}$/H is the abundance of the species
relative to hydrogen.
For the doubly ionized species, Ar$^{++}$ and Ne$^{++}$, which have three
fine-structure levels in their ground state, the emissivity is given by
\begin{equation}
j_{\rm III}=n_e n(q_{12}+q_{13})h\nu_{\rm III} {{\rm X_{III}}\over{\rm H}}=n_e n q_{\rm III}
h\nu_{\rm III} {{\rm X_{III}}\over{\rm H}},
\label{doubly_ionized}
\end{equation}
where $q_{12}$ and $q_{13}$ are the collisional excitation rates
between the first level and the second and third levels, respectively.
The \bra\ emissivity
\begin{equation}
j_{\rm Br\alpha}=\alpha_B(5\rightarrow 4) n_e n_{H^+} h\nu_{\rm Br\alpha},
\label{bracket_intensity}
\end{equation}
where $n_{H^+}$ is ionized hydrogen density $\alpha_B(5\rightarrow 4)$ is the case
B effective recombination coefficient for \bra. We assume $n\approx n_{H^+}$.

It follows that the abundance, X/H, is given by
\begin{equation}
{\rm X\over{H}}={{\rm X_{II}}\over{\rm H}}+{{\rm X_{III}}\over{\rm H}}={{\alpha(5\rightarrow 4)h\nu_{\rm Br\alpha}}\over{j_{\rm Br\alpha}}}\left[{{j_{\rm II}}\over{q_{\rm II}h\nu_{\rm II}}}+{{j_{\rm III}}\over{q_{\rm III}h\nu_{\rm III}}}\right].
\label{ne_abun_form}
\end{equation}

Inserting the basic atomic data assuming a gas temperature of $10^4$K one gets:
\begin{equation}
{{\rm Ne}\over{\rm H}}=10^{-5}\times{{2.30 f_{\rm Ne\ II}+f_{\rm Ne\ III}}\over{1.90 f_{\rm Br\alpha}}}\,\qquad {{\rm Ar}\over{\rm H}}=10^{-6}\times{{f_{\rm Ar\ II}+1.26 f_{\rm Ar\ III}}\over{1.43 f_{\rm Br\alpha}}},
\label{abun_eq}
\end{equation}
where $f_x$ are the line fluxes.

\begin{figure}[p]
\plotone{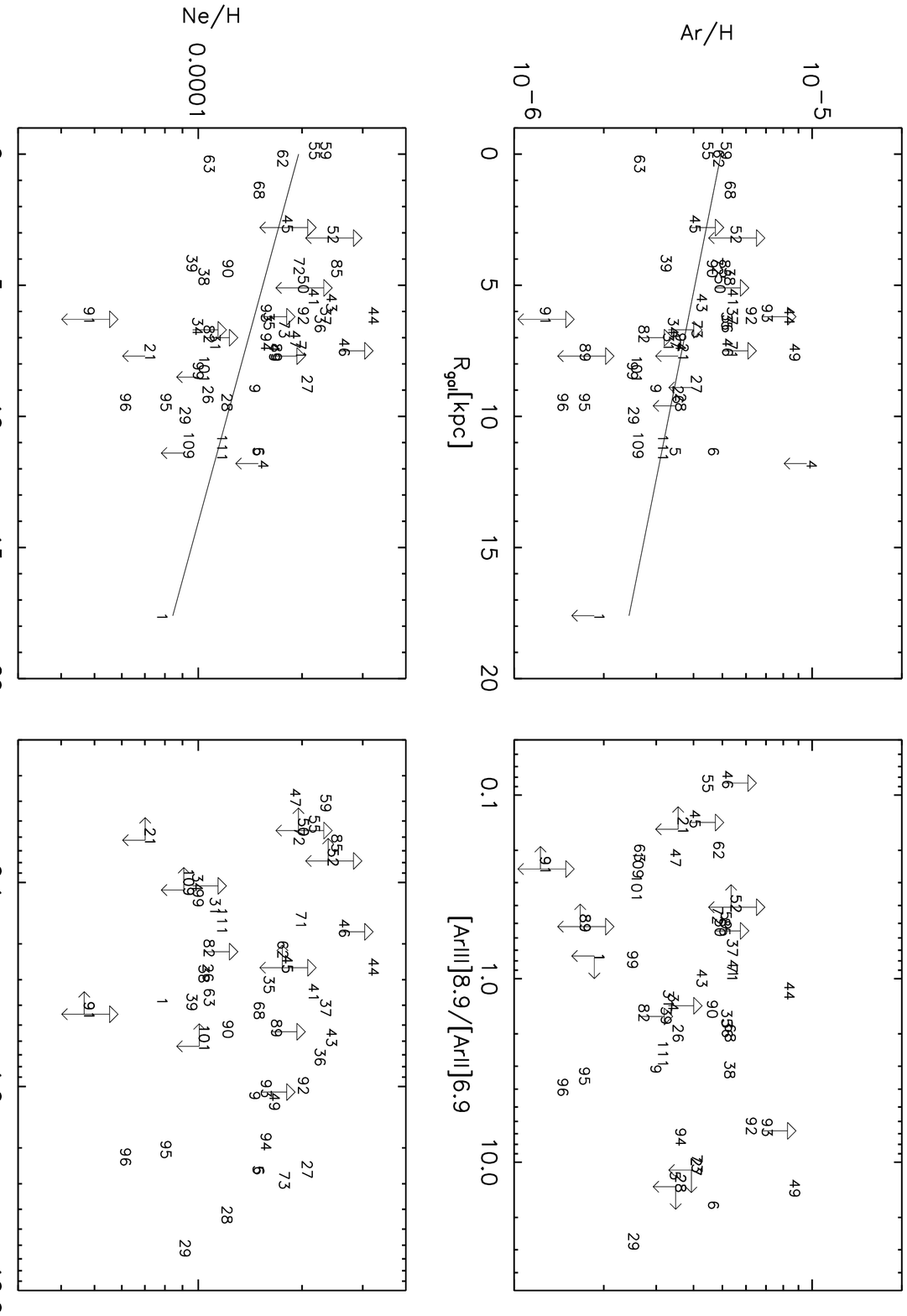}
\vspace{25cm}
\caption[Argon and neon abundances {\it
    vs.} galactocentric radius (left panels) and {\it vs.} the
  corresponding excitation ratio (right panels). Solid lines are fits to the
gradients (for $R_{gal}<$18 kpc), whose values are given in the text.]{}
\label{abundances}
\end{figure}

Figure~\ref{abundances} shows the computed neon and argon abundances {\it vs.}
galactocentric radius for the Galactic sources.
Both abundances show a decrease from
the inner parts of the Galaxy outwards. Both elements vary from $\sim 2\times$
solar to $\sim 1\times$ solar metallicities.
The fitted gradients over the entire range of galactocentric radii are
\begin{equation}
\left[{\rm Ne\over H}\right]=-3.71(\pm 0.06)-0.021(\pm 0.007)R_{gal},
\label{neon_abun_grad}
\end{equation}
and
\begin{equation}
\left[{\rm Ar\over H}\right]=-5.31(\pm 0.06)-0.018(\pm 0.008)R_{gal},
\label{argon_abun_grad}
\end{equation}
where the solar metallicity values are [Ne/H]$_{\odot}=-3.92$ and
[Ar/H]$_{\odot}=-5.60$ (Grevesse \& Sauval 1998).

The statistical significance of our detected gradients is low. However, our
results are consistent with previous studies. Afflerbach et al. (1997) found
similar gradients, from twice solar metallicity in
the inner Galaxy to lower values outwards, in their study of ${\rm O/H}$ and
${\rm S/H}$. Simpson et al. (1995) found a gradient
in the neon abundance which is steeper than ours ($-0.08\pm 0.02$ dex
kpc$^{-1}$), but their sample consists of only 18 sources. Simpson \&
Rubin (1990) also found the inner Galactic sources to be of about
twice solar metallicity. In their preliminary study Cox et al. (1999)
show evidence of a neon abundance gradient as probed by the
\neii/\bra\ ratio. However, their analysis
did not include contributions due to Ne$^{++}$ ions as probed by the
[Ne~{\small III}] emission which dominates in the high excitation nebulae. 

\section{Analysis}
\label{analysis}

The aim of the following analysis is to obtain a theoretical fit
to the empirical \linebreak \argon\ {\it vs.\/} \neon\ diagram shown in
Figure~\ref{ar_ne}.
We consider a sequence of photo-ionization models in
which we assume that the \hii regions are ionized by single stars.
We compute NLTE model atmospheres for high surface gravity dwarfs
and low gravity giants stars with a range of effective temperatures.
In the nebular analysis we explore the effects of
varying the gas density, ionization parameter and metallicity.
We also consider the effects of metallicity variations in the
model atmosphere computations.

\subsection{Models}
\label{models}

We use the {\it WM-basic} stellar atmosphere code\footnote{See also
  www.usm.uni-muenchen.de/people/adi/adi.html}
(Pauldrach et al. 2001) to compute the spectral energy distributions.
These NLTE atmosphere computations include the hydrodynamical
treatment of steady-state winds and mass loss, and include the combined effects
of line blocking and line blanketing. The {\it WM-basic} package employs
up-to-date atomic data, which provide the basis for a detailed
multi-level NLTE treatment of the metal ions, from C to Zn.
As discussed by Pauldrach et al., a key (and unique) feature of our model
atmospheres is that they successfully reproduce the spectral characteristics of
hot stars as revealed by high-resolution far-UV spectroscopy.
We use the NLTE atmospheres to construct model stars as input
ionizing sources for the nebulae which we model using the photo-ionization
code CLOUDY (version C90.04, Ferland et al. 1998)\footnote{See also
www.pa.uky.edu/$\sim$gary/cloudy/}.
We also carry out computations using the older generation LTE
static atmospheres (Kurucz 1979, Kurucz 1992).

\begin{figure}[p]
\plotone{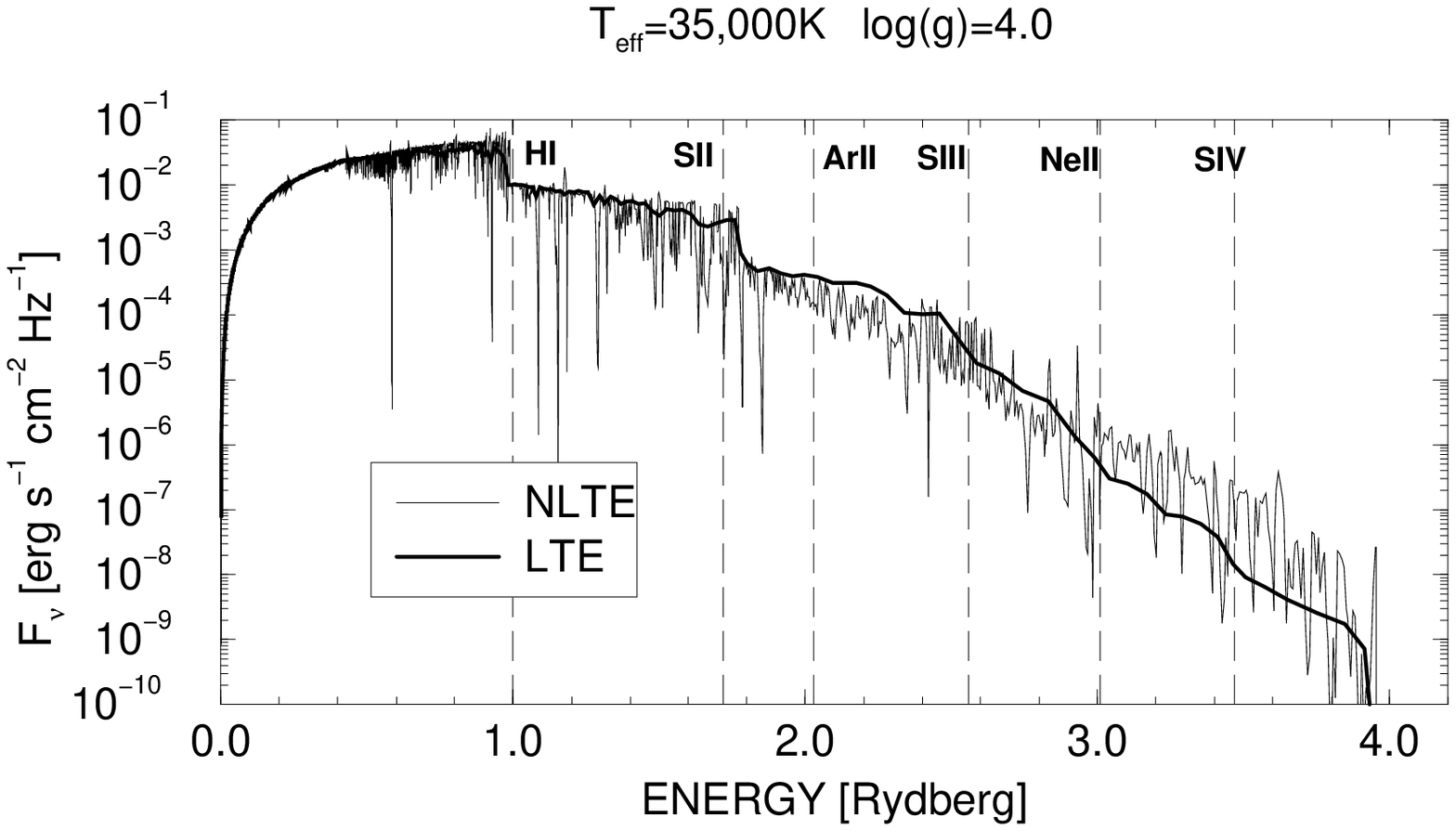}
\caption[Comparison between the LTE and NLTE SEDs, for $T_{eff}=35,000$K and $\log{g}=4.0$. The
ionization edges of the relevant ions are indicated by vertical dashed lines.]{}
\label{atmospheres}
\end{figure}

In general, the NLTE wind models have ``harder'' SEDs compared to
the hydrostatic LTE models (see also Schaerer \& de Koter 1997). 
For a given stellar effective temperature, the NLTE
atmospheres emit significantly larger fluxes of energetic photons.
Figure~\ref{atmospheres} shows an example of the differences between the
LTE and NLTE spectral energy distributions for an effective temperature 
$T_{eff}=35,000$K and surface gravity $\log{g}=4.0$.

The relative deficiency in high energy photons ($\gtorder 3$ Rydberg) in the
LTE model is clear. The ionization edges of the relevant ions are
overlayed on the spectra in this plot. The
nebular ionization state, expressed by the ratios \neon\ or \argon\, is
sensitive to the shape of the SED, and the NLTE models will generally
produce greater excitation. It is apparent from Figure~\ref{atmospheres},
that the difference in excitation produced by the
LTE and NLTE atmospheres will be more pronounced in the
Ne$^{++}$/Ne$^+$ ratio compared to Ar$^{++}$/Ar$^+$.
This helps in breaking the degeneracy between the SED shape and the ionization
parameter as factors which influence the ionization state of the gas.

We have computed model atmospheres for ``representative'' main sequence
dwarfs \linebreak ($\log{g}=4.0$) as well as super-giants ($\log{g}=3.25-3.75$)
for $T_{eff}$ ranging from 35,000 to 45,000 K, in steps of 10$^3$ K.
We computed models for solar ($1\ Z_{\odot}$; Grevesse \& Sauval 1998), 
half solar ($0.5\ Z_{\odot}$), and twice solar ($2\ Z_{\odot}$) metallicities. 
The stellar wind opacities and resulting SEDs depend on
the assumed stellar mass loss rates (Pauldrach et al. 2001).
For each metallicity we fix the mass-loss rates
(and terminal wind velocities) using the empirically based
wind momentum-luminosity relationship for O-star super-giants,
giants, and dwarfs (Kudritzki \& Puls 2000). 

We represent the nebulae as optically thick Str\"{o}mgren spheres, and consider
clouds with constant hydrogen densities ranging from
$10^2$ to $3\times 10^3$ cm$^{-3}$. We assume a gas filling factor $f=1$.
For each stellar effective
temperature, $T_{eff}$, we compute the Lyman continuum photon emission rate,
\begin{equation}
Q=4\pi
R_*^2\int^{\infty}_{\nu_0}{{F_{\nu}(T_{eff})}\over{h\nu}}d\nu\quad (s^{-1}),
\label{lyman_flux}
\end{equation}
where $F_{\nu}(T_{eff})$ (in erg s$^{-1}$ cm $^{-2}$ Hz$^{-1}$) is the SED,
$\nu_0$ is the Lyman limit frequency, and $R_*$ is the stellar radius
($\sim 10\ R_{\odot}$ for O-type dwarfs, and $\sim 20\ R_{\odot}$ for giants).

The nebular ionization structure depends on the shape of the SED, and also
on the ionization parameter
\begin{equation}
U\equiv{Q\over{4\pi r_s^2 nc}},
\label{ionization_parameter1}
\end{equation}
where $c$ is the speed of light, $n$ is the nebular gas density,
and $r_s$ is the Str\"{o}mgren radius given by
\begin{equation}
Q={{4\pi}\over 3}\alpha_B r_s^3 n^2,
\label{stromgren_radius}
\end{equation}
where $\alpha_B$ is the case B recombination coefficient.
It follows that
\begin{equation}
U={1\over c}\left({{n\alpha_B^2 Q}\over{36\pi}}\right)^{1/3} \ \ \ .
\label{ionization_parameter}
\end{equation}
Thus, for a fixed nebular density the ionization
parameter is fully determined by the stellar atmosphere model. 
For a given density we determine $U$ from the stellar model, 
and compute the resulting \argon\ and \neon\ emission line ratios using CLOUDY.
As a representative density we set $n=800$ cm$^{-3}$ as determined
from the S~{\small III} density diagnostic (see \S 3.3). 

Tables~\ref{dwarfs_models}, \ref{giants_models}, and~\ref{LTE_models} summarize
the parameters for the stellar atmospheres
we adopt as the ionizing sources in our photoionization computations. These
parameters include the stellar effective temperature, surface gravity, stellar
radius, wind parameters (mass-loss rate $\dot{M}$ and the terminal velocity
$v_{\infty}$), and hydrogen photoionizion rates for NLTE typical dwarfs and 
super-giants, and for LTE dwarfs. The parameters are given for the half-solar,
solar, and twice-solar metallicities.

\subsection{Comparison with Observations}
\label{comparison}

\begin{figure}[p]
\plotone{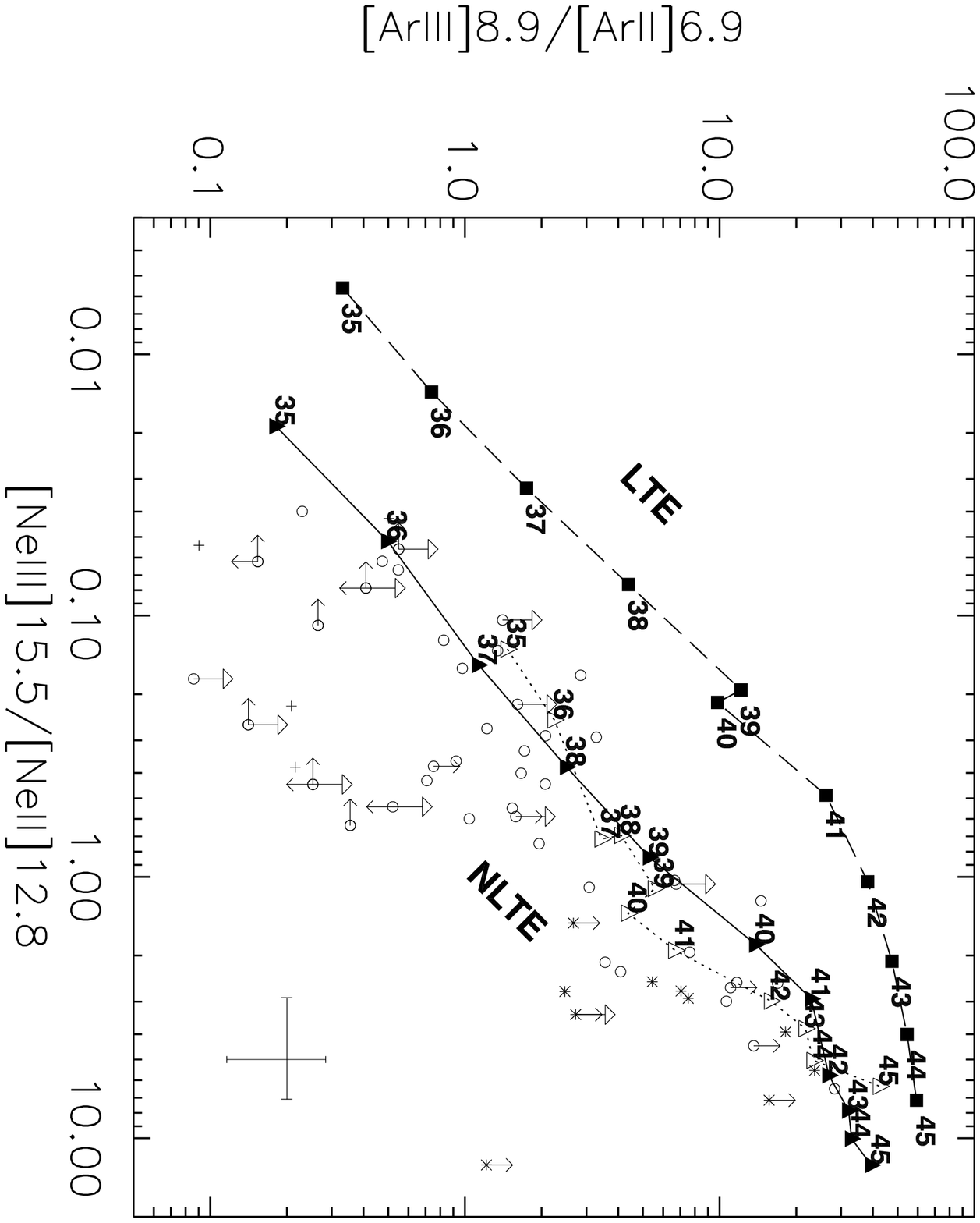}
\vspace{25cm}
\caption[\argon\ {\it vs.\/} \neon\ diagnostic diagram. Bold faced
numbers indicate models. Triangles are
  1 $Z_{\odot}$ NLTE models. The dwarfs sequence is designated by solid line,
and the super-giants sequence is designated by dotted line. Squares are 1
$Z_{\odot}$ LTE models. Models are designated by their effective temperature in
  $10^3$K. A representative error bar
for the data is shown in the lower right corner.]{}
\label{ar_ne_models}
\end{figure}

In Figure~\ref{ar_ne_models} we show the \argon\ {\it vs.\/} \neon\
observations (Figure~\ref{ar_ne}) together with the model computations
for the LTE (dwarfs) and NLTE (dwarfs and giants) stars for
solar metallicity, and a gas density of 800 cm$^{-3}$.
The bold faced numbers beside each model point give the $T_{eff}$
of the corresponding star in units of $10^3$ K. A representative error bar
for the data is shown in the lower right corner of the plot.

For all sequences the predicted nebular excitation increases with increasing
effective temperature of the photoionizing star. However, the
LTE and NLTE models are clearly offset from one another, and it is evident that
the NLTE sequences provide a much better fit to the observed
excitation correlation. For a given \argon\ ratio the 
LTE models predict \neon\ ratios which are factors of 
$\sim 10$ smaller than observed. This discrepancy appears to be
almost entirely eliminated in the NLTE sequences.
The NLTE fit may actually be even better than is indicated
by Figure~\ref{ar_ne_models} since some of the scatter is likely due to
underestimated extinction corrections for the \ariii\ line.

\begin{figure}[p]
\plotone{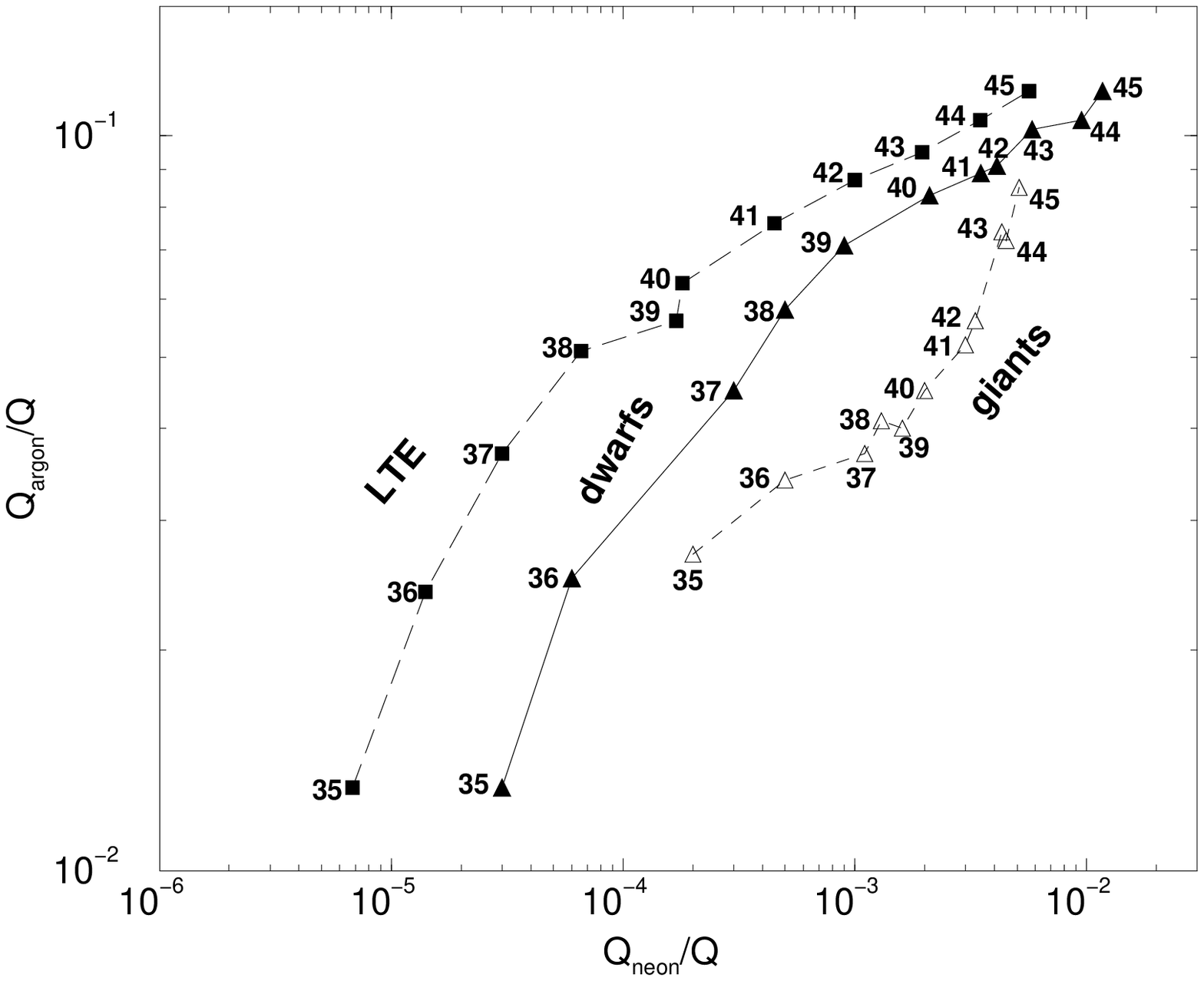}
\vspace{25cm}
\caption[The fraction of photons that can ionize Ar$^{+}$, $Q_{argon}/Q$ {\it
vs.\/} the fraction for Ne$^{+}$, $Q_{neon}/Q$, for each one of the model SEDs.
The plot follows exactly the behavior of the models in the excitation ratios
diagram, Figure~\ref{ar_ne_models}.]{}
\label{ar_ne_fint}
\end{figure}

The key difference between the LTE and NLTE models lies
in the different spectral shapes of their photoionizing continua.
This is illustrated in Figure~\ref{ar_ne_fint} where
for each effective temperature in the LTE and NLTE sequences 
we plot the fractions $Q_{argon}/Q$ {\it vs.\/} $Q_{neon}/Q$,
where $Q_{argon}$ is the photon emission rate for photons above
the Ar$^+$ 27.6 eV ionization threshold, $Q_{neon}$ is the emission
rate above the Ne$^+$ 40.96 eV threshold, and $Q$ is the total
Lyman continuum (Lyc) emission rate.  The SEDs harden as $T_{eff}$ increases,
and both ratios $Q_{argon}/Q$ and $Q_{neon}/Q$ increase.
For a given value of $Q_{argon}/Q$
the LTE atmosphere models produce much smaller
values of $Q_{neon}/Q$
than the NLTE atmospheres. This is the behavior that is 
essentially represented in the \argon\ {\it vs.\/} \neon\
excitation diagram. The ISO fine-structure data thus provide
direct empirical evidence for the improved accuracy of the
NLTE atmosphere SEDs in the Lyman continuum range.
The NLTE sequence also indicates that for most of the
displayed \hii regions the effective temperatures of the exciting
stars plausibly lie in the range of 35,000 to 45,000 K.

We have examined the dependence of the model results
on the ionization parameter by considering variations
in both $Q$ and $n$ (see eq.~\ref{ionization_parameter}). 
Although $Q$ is fixed by the stellar model, we have verified that
changing $Q$ (for a fixed SED) simply moves the computed points
along the (approximate) diagonal lines defined
by each sequence of models. Larger values of $Q$ could
be produced by clusters of (identical) stars of given
effective temperature. For an \hii region excited by a mix of stars one would
expect the high SED energies to be dominated by hotter stars 
and the low energies by cooler stars.
We will explore the effects of realistic clusters containing a synthesized
distribution of stars and associated cluster SEDs in a later paper.
Variations in the assumed density (in the range
of 10$^2$ to $3\times 10^3$ cm$^{-3}$) induce only 
small shifts in the positions of the computed model sequences. 

\begin{figure}[p]
\plotone{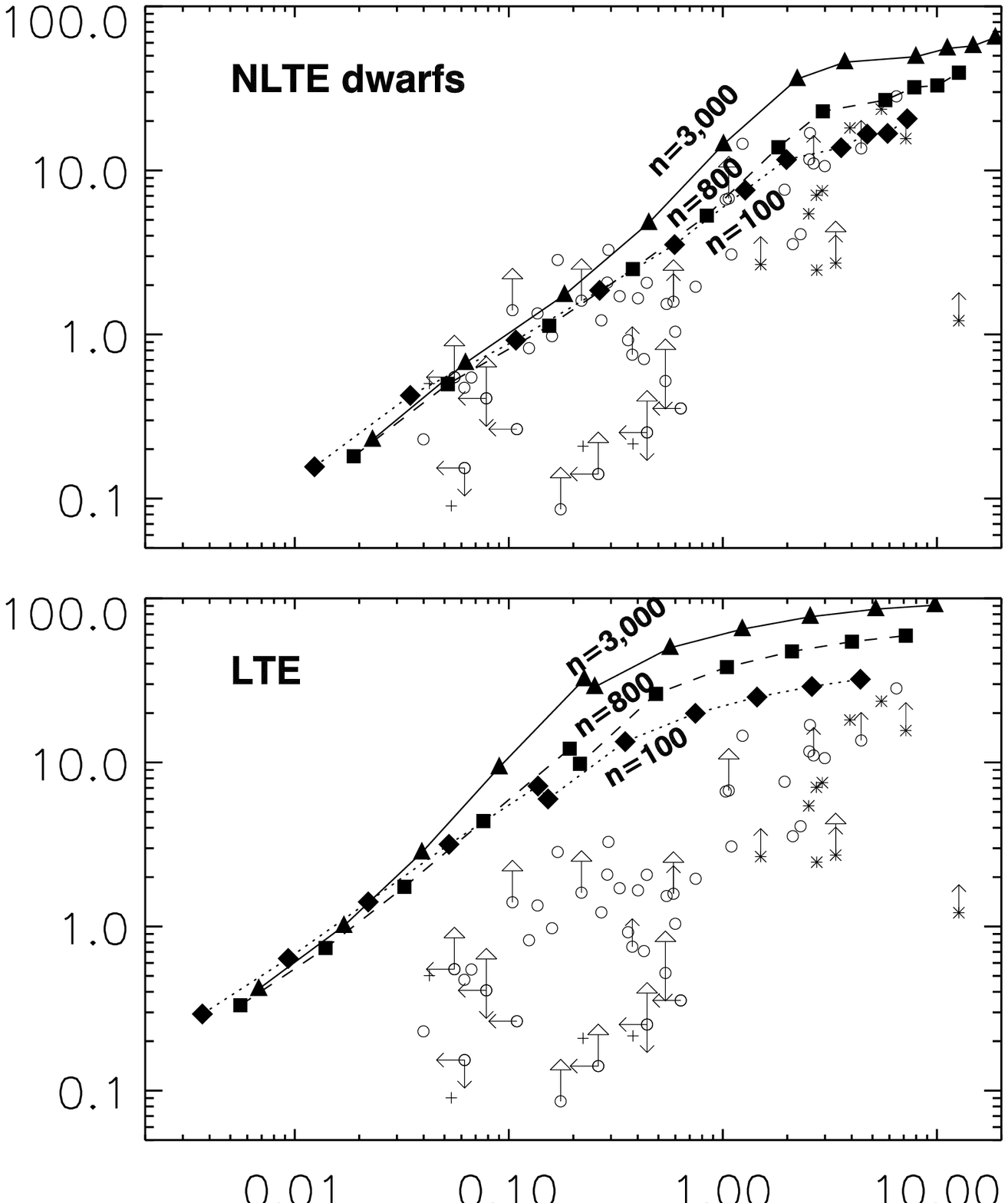}
\vspace{25cm}
\caption[\argon\ {\it vs.\/} \neon\ diagnostic diagram with the NLTE
dwarfs models (upper panel) and the LTE models (lower panel) of
various gas densities over-plotted. Triangles are
  3,000 cm$^{-3}$ models. Squares are 800 cm$^{-3}$ models, and
  diamonds are 100 cm$^{-3}$ models. All model sequences are composed
  of models with effective temperatures in the range 35 (bottom left)
  to 45 (top right) $\times 10^3$K.]{}
\label{ar_ne_den}
\end{figure}

In Figure~\ref{ar_ne_den} we show the NLTE dwarfs and LTE model sequences for
nebular densities of 100, 800, and $3\times 10^3$ cm$^{-3}$.
We conclude that the LTE sequence cannot be reconciled with the
data by varying either $Q$ or $n$.

\begin{figure}[p]
\plotone{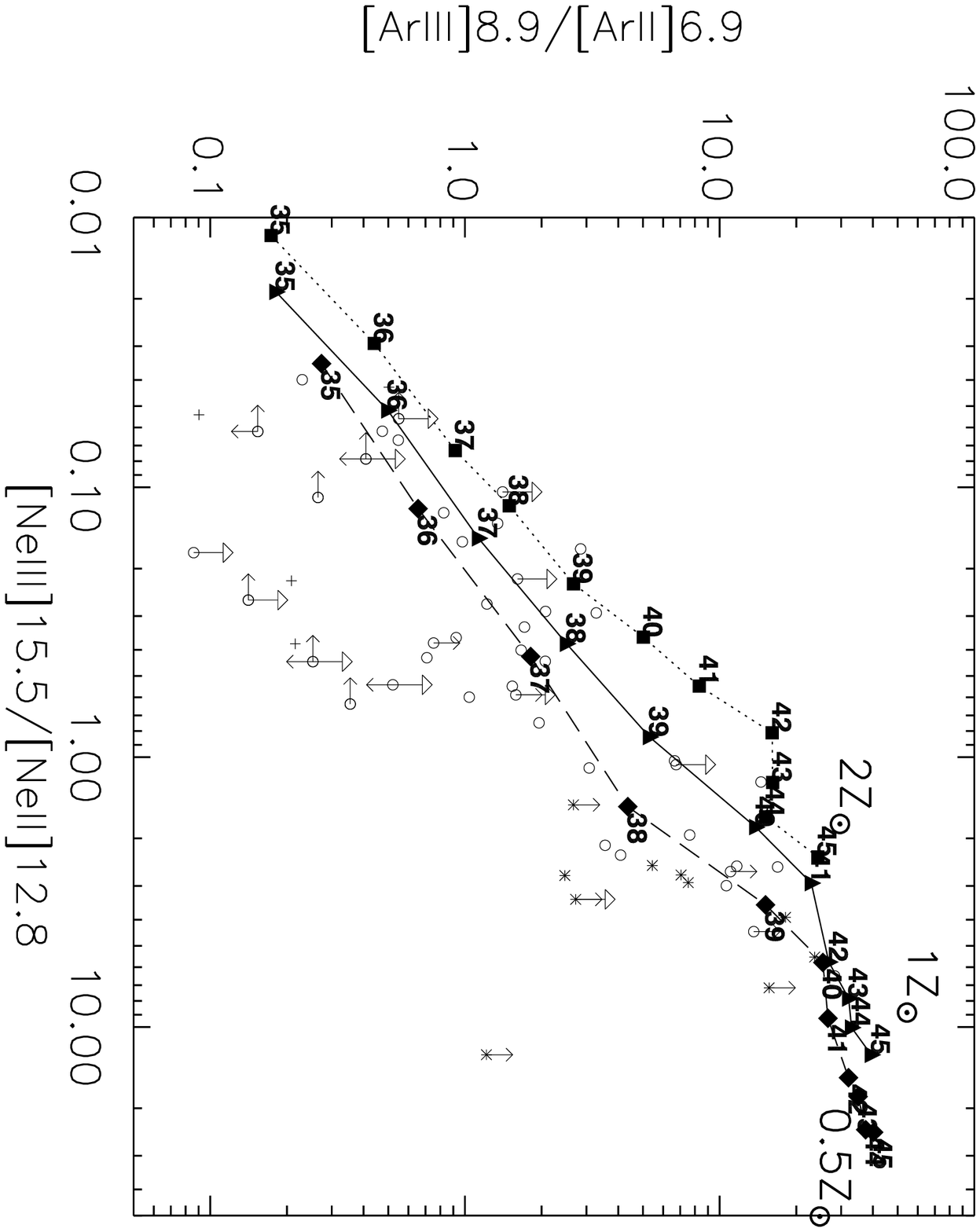}
\vspace{25cm}
\caption[\argon\ {\it vs.\/} \neon\ diagnostic diagram for dwarfs models with
various metallicities. Triangles are
  solar metallicity models. Diamonds and squares indicate
  0.5 $Z_{\odot}$ and 2 $Z_{\odot}$, respectively. Models are
  designated by their effective temperature in $10^3$K.]{}
\label{ar_ne_metal}
\end{figure}

In order to investigate the influence of varying the metallicity 
we show in Figure~\ref{ar_ne_metal} results for three sequences of NLTE
atmosphere models of dwarfs -- computed assuming half-solar, solar, and
twice-solar metallicities.
We varied the metallicity self-consistently in the nebulae and in the stellar
atmospheres. Both of these variations affects the computed line
ratios. Lower nebular metallicity results in higher gas temperatures,
reduced recombination rates, and therefore slightly higher excitation.
Afflerbach et al. (1997) found a similar trend in their study,
but they also
stressed the need for exploring the effects of metallicity variations
in the stellar atmospheres as well. We have done this here,
and find that such variations have a somewhat larger effect on
the computed line ratios.
For a given effective temperature the SED hardens with
decreasing metallicity, due to decreased line blocking in the 
stellar atmosphere winds.  This effect is primarily significant
above $\sim 2.5$ Ryd.  At these energies
the blocking is dominated by many Fe~{\small IV}, Fe~{\small V} and
Fe~{\small VI} lines, which are 
mainly optically thick near the wind sonic point and optically thin in the 
outer wind region. As the metallicity is increased, the Fe lines remain 
optically thick to higher velocities, thus enhancing the blocking influence 
in this frequency regime. However, 
below $\sim 2.5$ Ryd the blocking is dominated by 
C, N, O, Ne, and Ar. The corresponding lines are mainly optically thick in the 
entire wind region. Changes in metallicity therefore do not significantly
affect the emergent (blocked) fluxes below 2.5 Ryd.
Therefore, varying the metallicity has a larger effect on the
\neon\ ratio than on the \argon\ ratio as indicated in
Figure~\ref{ar_ne_metal}.
For example, for a 40,000 K star increasing the nebular and
stellar metallicty from 0.5 to 2 $Z_\odot$ reduces the
\argon\ ratio by a factor of $\sim 5$ and the \neon\ ratio
by a factor of $\sim 20$.

Figure~\ref{ar_ne_metal} also demonstrates that the excitation in the LMC
nebulae (indicated by star symbols) is plausibly enhanced, at least in part,
by the reduced metallicity of the LMC.
However, the low-excitation of the inner Galaxy sources (indicated by the
crosses) is probably {\it not} due to their large metallicities,
since these objects appear to have relatively large
\neon\ ratios for their \argon\ ratios, whereas
increasing the metallicity should selectively 
{\it reduce} the \neon\ ratio. The low excitation
of the inner Galaxy sources is most likely due to
lower effective temperatures of the stellar (or
cluster) sources. In this respect these objects
may resemble the typical low-excitation sources
observed in the starburst sample of Thornley et al. (2000).

\section{Discussion and Summary}
\label{conclusions}

We have not attempted detailed modeling of individual sources
in this paper. However, as we have shown, the ensemble of
Galactic fine-structure emission line data provide empirical
support for the generally harder photoionizing SEDs predicted
in NLTE wind driven line blocked models of hot-star atmospheres.
In particular, such models can readily
account for the presence of high excitation
[Ne~{\small III}] emission lines in nebular spectra. Earlier studies
which had relied on LTE atmospheres in the nebular analysis
were unable to account for the observed [Ne~{\small III}] emission,
and under-predicted the line intensities by factors of $\sim 10$
(e.g. Baldwin et al. 1991; Rubin et al. 1995; Simpson et al. 1995).
Our study largely resolves this issue
and supports the suggestions of Sellmaier et al. (1996) and Stasinska \&
Schaerer (1997) that this ``Ne~{\small III} problem'' was due primarily to a
significant under-prediction of Lyc photons above $\sim 40$ eV in the
Kurucz LTE atmospheres.  

This conclusion has important consequences for the interpretation
of extragalactic fine-structure line observations. 
Thornley et al. (2000) carried out detailed starburst modeling
of the \neon\ ratio expected from \hii regions ionized by
clusters of stars in which the stellar SEDs were modeled
using the new generation NLTE atmospheres. Given that 
the hottest stars in such models easily produce large
nebular \neon\ ratios, the low ratios actually observed
led to the conclusion that the relative number of
hot stars is small due to aging of
the starburst systems. Our analysis supports the conclusion
that low \neon\ ratios require the removal of the hottest
stars as dominant contributors to the ionization of
the starburst galaxies. 

We summarize our key results.
We have compiled
archival mid-IR ISO-SWS fine-structure lines and \hii recombination
line observations from \hii regions in the Galaxy, the LMC, SMC, and M33.
We find that:
\begin{itemize}
\item[1.] There is a steep outward increase in excitation in the
  Galaxy as probed by the fine-structure line ratios
\neon, \argon, and \sulfur. Sources in the central kpc of the
  Galaxy show the lowest levels of excitation, and the LMC, and SMC
  sources show the highest excitation level.
\item[2.] There is a slight decrease in metallicity from $\sim 2Z_{\odot}$ in
the Galactic center to $\sim 1Z_{\odot}$ in the outer disk.
\item[3.] The increase in nebular excitation across
the Galactic disk is likely due mainly to a
rise in the typical effective temperatures of the exciting stars
rather than a hardening of the stellar SEDs and enhanced
nebular excitation associated with the decline in metallicity
at fixed effective temperature. 
\item[4.] By comparing models to observations, we find that the NLTE wind
  model atmospheres represent an impressive improvement in the realistic
  description of hot star ionizing spectra (compared to LTE static models).
  In particular the [Ne~{\small III}] emission observed in Galactic \hii
regions is readily accounted for by these atmospheres.
\item[5.] We estimate an ``integrated'' \neon\ ratio of 0.8
for the Galaxy, well inside the range of values found for
starburst galaxies. The good fit of the NLTE models to the data, support the
suggestion by Thornley et al. (2000) that the observed \neon\ ratio in
starburst galaxies is small due to aging effects.
\end{itemize}

\section*{Acknowledgments}

We thank Roland Tacke and Julia Morfill for help with the data
reduction, and Tadziu Hoffmann for his assistance with the
{\it WM-basic} program.
We thank Gary Ferland and Hagai Netzer for discussions, and
the anonymous referee for helpful suggestions.
The ISO spectrometer data center at MPE is supported by DLR under
grants 50 QI 8610 8 and 50 QI 9402 3. Our research is supported by the
German-Israeli Foundation (grant I-0551-186.07/97).
\pagebreak

\listoffigures

\newpage

\begin{table}[p]
\caption{IR fine-structure lines, which were measured in the spectra
  selected from the ISO archive, with their wavelengths, ionization
  potentials, and critical densities.}
\vspace{0.5cm}
\begin{tabular}[h]{cccc} \hline\hline
Line & Wavelength & Ionization Potential & Critical Density \\
     &  ($\mu$m)  & (eV) & (cm$^{-3}$) \\ \hline
\bra &  4.05 & & \\
\brb &  2.63 & & \\
$[$Ar~{\small II}$]$ &  6.99 & 27.63 & 4.17$\times 10^5$ \\
$[$Ar~{\small III}$]$&  8.99 & 40.74 & 2.60$\times 10^5$ \\
$[$Ne~{\small II}$]$ & 12.81 & 40.96 & 7.00$\times 10^5$ \\
$[$Ne~{\small III}$]$& 15.56 & 63.45 & 2.68$\times 10^5$ \\
$[$Ne~{\small III}$]$& 36.01 & 63.45 & 5.50$\times 10^4$ \\
$[$S~{\small III}$]$ & 18.71 & 34.79 & 2.22$\times 10^4$ \\
$[$S~{\small III}$]$ & 33.48 & 34.79 & 7.04$\times 10^3$ \\
$[$S~{\small IV}$]$  & 10.51 & 47.22 & 5.39$\times 10^4$ \\
\hline
\end{tabular}
\label{linelist}
\end{table}

\clearpage

\begin{figure*}
\includegraphics{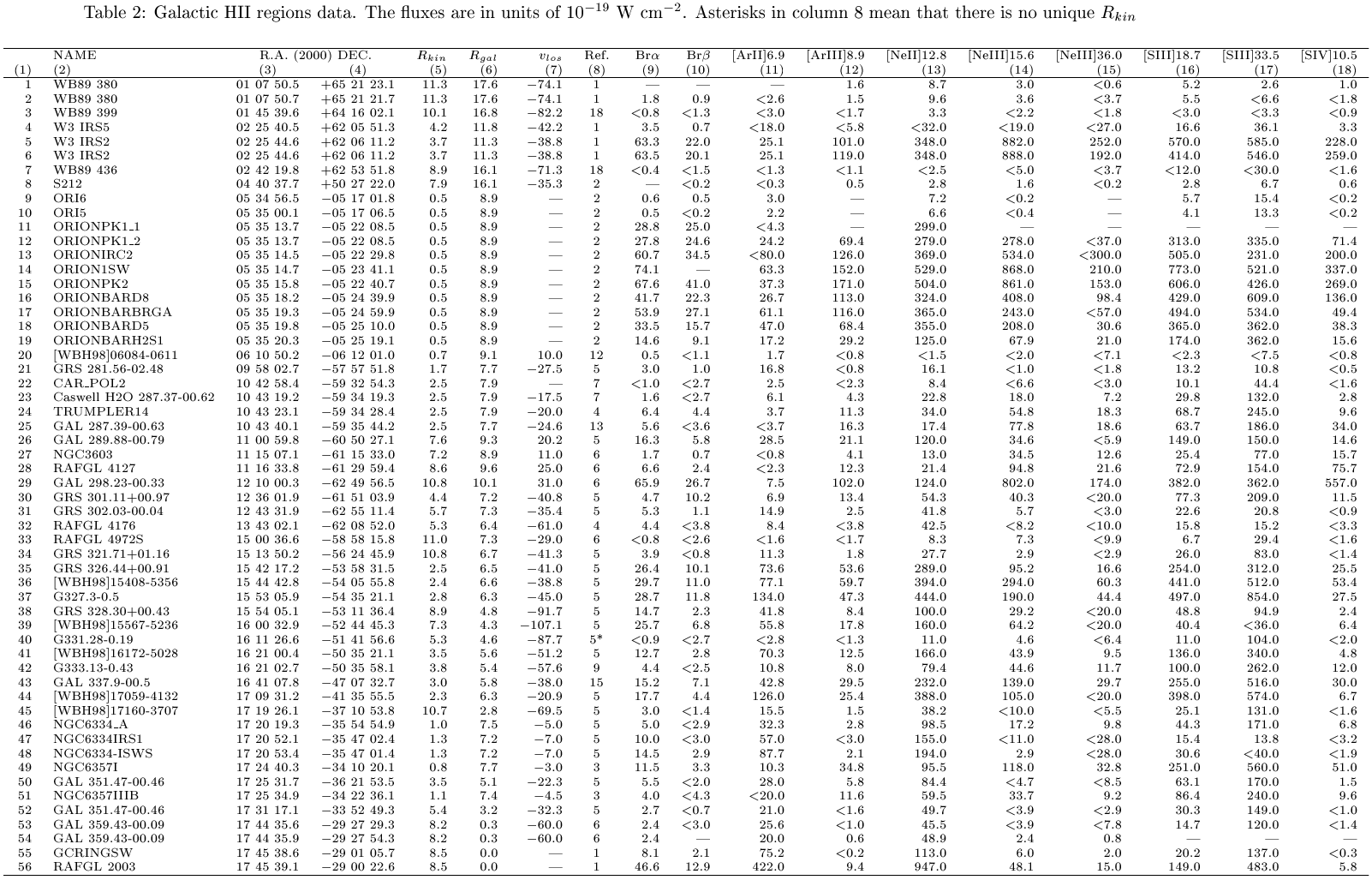}
\vspace{25cm}
\end{figure*}

\begin{figure*}
\includegraphics{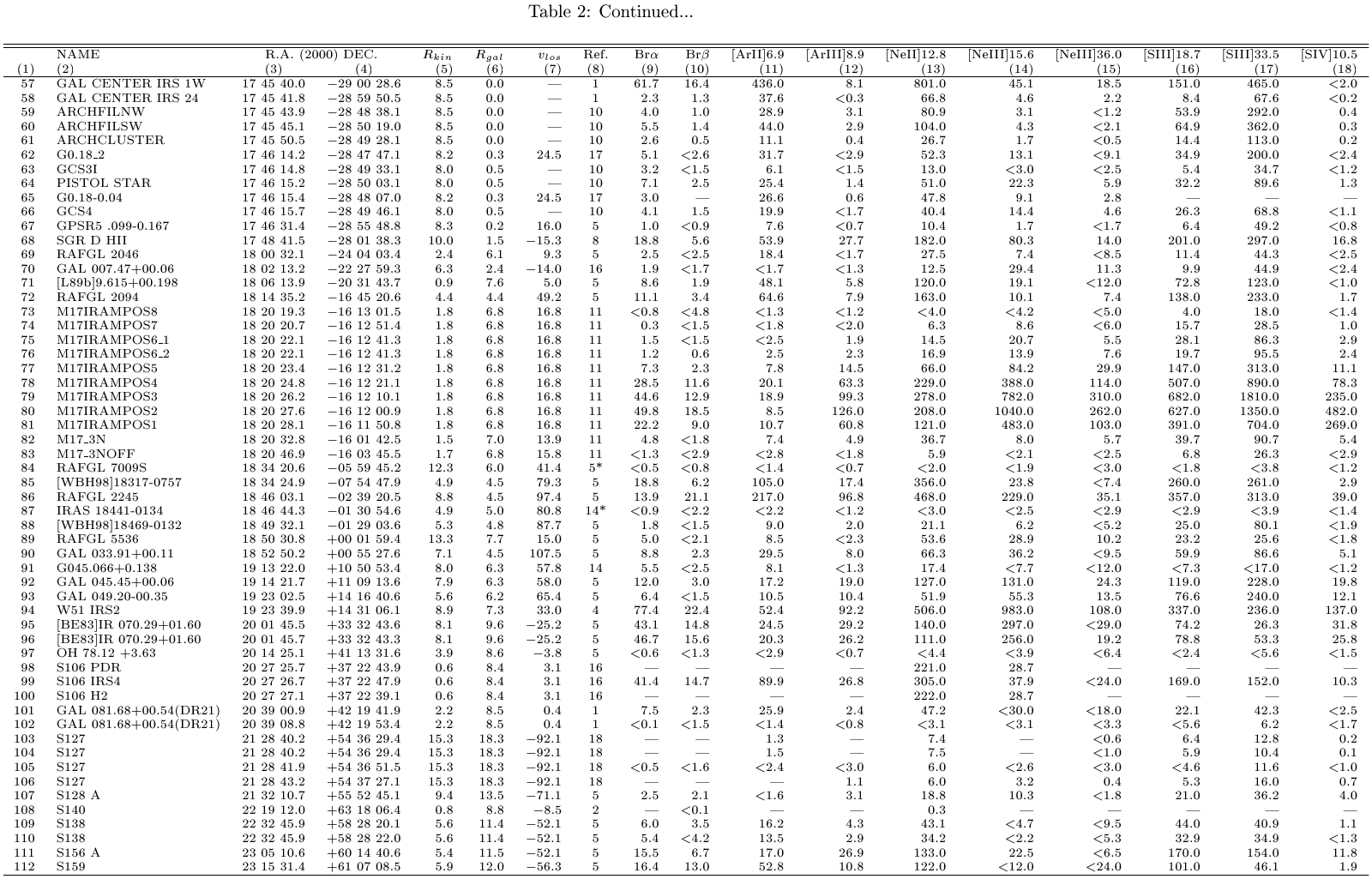}
\vspace{25cm}
\end{figure*}
\addtocounter{table}{1}

\clearpage

\begin{table}[p]
\caption{Abbreviation for the references of Tables 2.}
\vspace{0.5cm}
\begin{tabular}{lcl}\hline\hline
1-Afflerbach et al. 1996         & & 10-Lang et al. 1999 \\
2-Blitz, Fich, \& Stark 1982     & & 11-Lockman 1989\\
3-Brand et al. 1984              & & 12-Shepherd \& Churchwell 1996 \\
4-Braz \& Epchtein 1983          & & 13-Slysh et al. 1994 \\
5-Bronfman, Nyman, \& May 1996   & & 14-Szymczak, Hrynek, \& Kus 2000 \\
6-Caswell \& Haynes 1987         & & 15-Walsh et al. 1997 \\
7-Caswell et al. 1989            & & 16-Wink, Altenhoff, \& Mezger 1982 \\
8-Churchwell, Walmsley, \& Cesaroni 1990& & 17-Wink, Wilson, \& Bieging 1983 \\
9-Huang et al. 1999              & & 18-Wouterloot \& Brand 1989 \\
\hline
\end{tabular}
\label{data_ref}
\end{table}

\begin{table}[p]
\caption{Complex regions, whose intensities were summed up.}
\vspace{0.5cm}
\begin{tabular}{lc}\hline\hline
OBJECT                                 & OBSERVATIONS \\
                                       & (From Table 2) \\ \hline
Arches Cluster                         & 59--61 \\
Carina Nebula                          & 22--25 \\
IRAS 17431-2846                        & 62, 65 \\
IRAS 16172-5134                        & 41, 42 \\
IRAS 17424-2859 (Sagittarius A IRS 24) & 55--58 \\
IRAS 17430-2848 (Pistol star)          & 63,64,66 \\
Omega Nebula (M17)                     & 73--81 \\
NGC 6334                               & 47, 48 \\
Orion Nebula (M42)                     &  9--19 \\
GAL 359.43-00.09                       & 53, 54 \\
\hline
\end{tabular}
\label{multiple_obs}
\end{table}

\begin{figure*}
\includegraphics{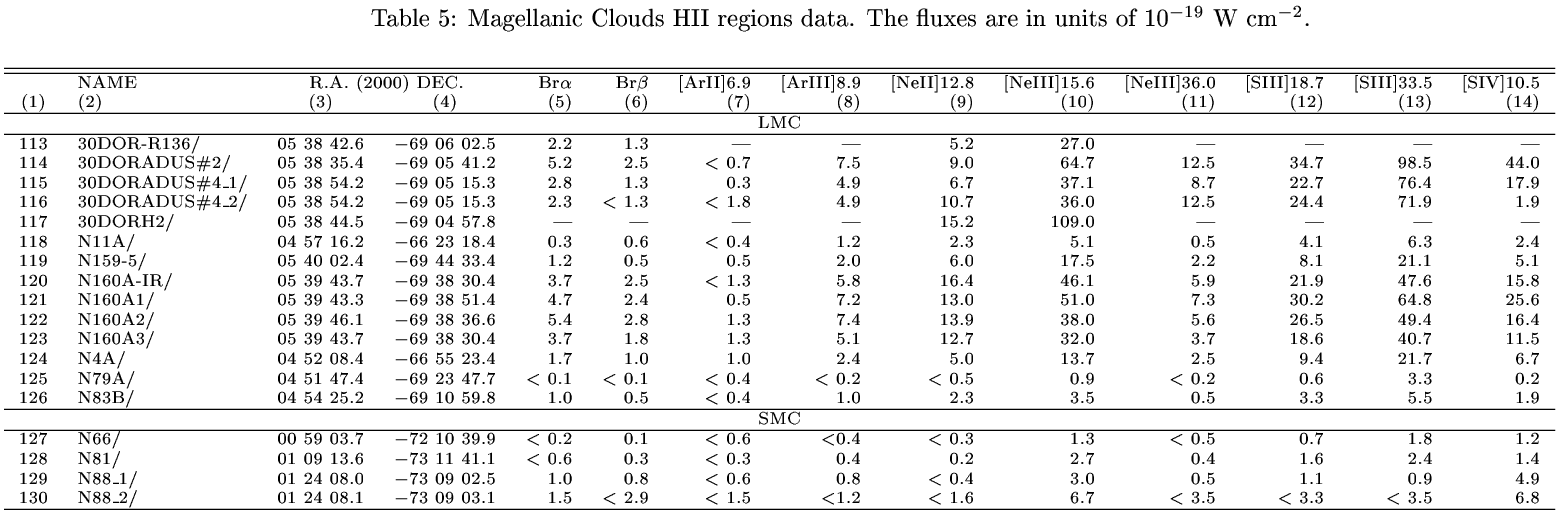}
\vspace{25cm}
\end{figure*}
\addtocounter{table}{1}

\begin{figure*}
\includegraphics{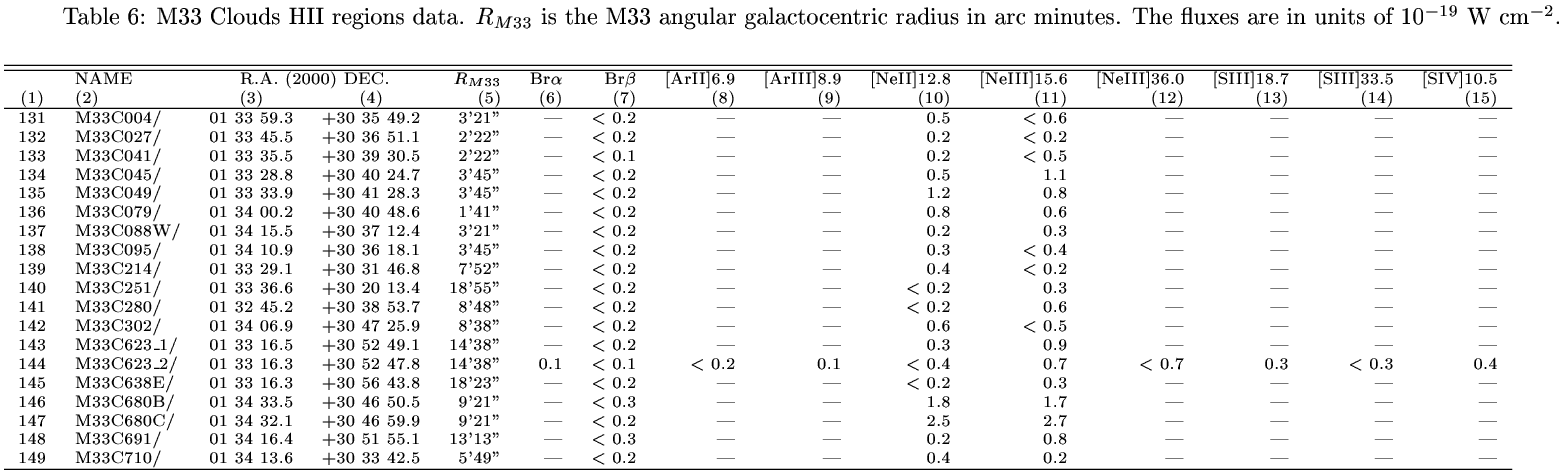}
\vspace{25cm}
\end{figure*}
\addtocounter{table}{1}

\begin{table}[p]
\caption{The model atmospheres parameters -- stellar effective temperature,
 surface gravity, stellar radius, wind parameters ($\dot{M}$ and $v_{\infty}$)
and hydrogen photoionizion rates of NLTE typical dwarfs. The parameters are
given for half-solar, solar, and twice-solar metallicities.}
\vspace{0.5cm}
\scriptsize
\begin{tabular}{cccccccccccc}\hline\hline
 & & & \multicolumn{3}{c}{$0.5 Z_{\odot}$} & \multicolumn{3}{c}{$1 Z_{\odot}$} & \multicolumn{3}{c}{$2 Z_{\odot}$} \\ \hline

 $T_{eff}$ & $\log{g}$ & $R_{\star}$ & $\log{Q}$ & $v_{\infty}$ & $10^{-7}\dot{M}$ & $\log{Q}$ & $v_{\infty}$ & $10^{-7}\dot{M}$ & $\log{Q}$ & $v_{\infty}$ & $10^{-7}\dot{M}$ \\ 

 (K) & ($cm\ s^{-2}$) & ($R_{\odot}$) & ($s^{-1}$) & ($km s^{-1}$) & ($M_{\odot} yr^{-1}$) & ($s^{-1}$) & ($km s^{-1}$) & ($M_{\odot} yr^{-1}$) & ($s^{-1}$) & ($km s^{-1}$) & ($M_{\odot} yr^{-1}$) \\ \hline

 35,000 & 4.0 &  9.0 & 48.37 & 2000 & 0.8 & 48.35 & 2004 & 1.6 & 48.37 & 2008 &  3.4 \\
 36,000 & 4.0 &  9.2 & 48.52 & 2079 & 1.0 & 48.51 & 2093 & 1.9 & 48.53 & 2086 &  3.9 \\
 37,000 & 4.0 &  9.4 & 48.67 & 2158 & 1.2 & 48.68 & 2177 & 2.3 & 48.68 & 2181 &  4.6 \\
 38,000 & 4.0 &  9.6 & 48.76 & 2283 & 1.5 & 48.73 & 2272 & 2.8 & 48.73 & 2277 &  5.2 \\
 39,000 & 4.0 &  9.8 & 48.88 & 2358 & 1.8 & 48.87 & 2362 & 3.3 & 48.85 & 2365 &  6.1 \\
 40,000 & 4.0 & 10.0 & 48.99 & 2450 & 2.2 & 48.98 & 2450 & 4.0 & 48.98 & 2460 &  7.0 \\
 41,000 & 4.0 & 10.2 & 49.09 & 2548 & 2.8 & 49.08 & 2541 & 4.8 & 49.07 & 2564 &  8.3 \\
 42,000 & 4.0 & 10.4 & 49.19 & 2637 & 3.4 & 49.18 & 2640 & 5.7 & 49.16 & 2620 &  9.6 \\
 43,000 & 4.0 & 10.6 & 49.26 & 2732 & 4.0 & 49.26 & 2719 & 6.7 & 49.26 & 2722 & 11.0 \\
 44,000 & 4.0 & 10.8 & 49.34 & 2803 & 4.9 & 49.33 & 2806 & 7.8 & 49.32 & 2800 & 13.0 \\
 45,000 & 4.0 & 11.0 & 49.42 & 2894 & 5.8 & 49.40 & 2923 & 9.1 & 49.39 & 2900 & 14.0 \\
\hline
\end{tabular}
\label{dwarfs_models}
\end{table}

\begin{table}[p]
\caption{The model atmospheres parameters of our NLTE typical giants for
half-solar, solar, and twice-solar metallicities.}
\vspace{0.5cm}
\scriptsize
\begin{tabular}{cccccccccccc}\hline\hline
 & & & \multicolumn{3}{c}{$0.5 Z_{\odot}$} & \multicolumn{3}{c}{$1 Z_{\odot}$} & \multicolumn{3}{c}{$2 Z_{\odot}$} \\ \hline

 $T_{eff}$ & $\log{g}$ & $R_{\star}$ & $\log{Q}$ & $v_{\infty}$ & $10^{-7}\dot{M}$ & $\log{Q}$ & $v_{\infty}$ & $10^{-7}\dot{M}$ & $\log{Q}$ & $v_{\infty}$ & $10^{-7}\dot{M}$ \\ 

 (K) & ($cm\ s^{-2}$) & ($R_{\odot}$) & ($s^{-1}$) & ($km s^{-1}$) & ($M_{\odot} yr^{-1}$) & ($s^{-1}$) & ($km s^{-1}$) & ($M_{\odot} yr^{-1}$) & ($s^{-1}$) & ($km s^{-1}$) & ($M_{\odot} yr^{-1}$) \\ \hline

 35,000 & 3.25 & 24.0 & 49.54 & 1988 & 48 & 49.54 & 1998 &  63 & 49.52 & 2008 &  83 \\
 36,000 & 3.30 & 23.4 & 49.60 & 2058 & 52 & 49.60 & 2061 &  69 & 49.59 & 2083 &  91 \\
 37,000 & 3.35 & 22.8 & 49.64 & 2122 & 54 & 49.63 & 2125 &  74 & 49.62 & 2128 &  99 \\
 38,000 & 3.40 & 22.2 & 49.65 & 2172 & 58 & 49.67 & 2180 &  78 & 49.66 & 2193 & 110 \\
 39,000 & 3.45 & 21.6 & 49.72 & 2227 & 61 & 49.69 & 2235 &  83 & 49.70 & 2240 & 110 \\
 40,000 & 3.50 & 21.0 & 49.74 & 2313 & 64 & 49.75 & 2311 &  88 & 49.74 & 2329 & 120 \\
 41,000 & 3.55 & 20.4 & 49.79 & 2364 & 66 & 49.76 & 2367 &  91 & 49.77 & 2376 & 130 \\
 42,000 & 3.60 & 19.8 & 49.81 & 2424 & 68 & 49.78 & 2426 &  95 & 49.79 & 2431 & 140 \\
 43,000 & 3.65 & 19.2 & 49.84 & 2493 & 70 & 49.85 & 2483 & 100 & 49.82 & 2490 & 140 \\
 44,000 & 3.70 & 18.6 & 49.86 & 2556 & 71 & 49.86 & 2548 & 100 & 49.85 & 2569 & 150 \\
 45,000 & 3.75 & 18.0 & 49.87 & 2609 & 73 & 49.88 & 2610 & 110 & 49.87 & 2618 & 160 \\
\hline
\end{tabular}
\label{giants_models}
\end{table}

\begin{table}[p]
\caption{The model atmospheres parameters of the LTE dwarfs for solar
metallicity.}
\vspace{0.5cm}
\scriptsize
\begin{tabular}{cccc}\hline\hline
 $T_{eff}$ & $\log{g}$ & $R_{\star}$ & $\log{Q}$ \\

 (K) & ($cm\ s^{-2}$) & ($R_{\odot}$) & ($s^{-1}$) \\ \hline

 35,000 & 4.0 &  9.0 & 48.43 \\
 36,000 & 4.0 &  9.2 & 48.57 \\
 37,000 & 4.0 &  9.4 & 48.70 \\
 38,000 & 4.0 &  9.6 & 48.81 \\
 39,000 & 4.0 &  9.8 & 48.92 \\
 40,000 & 4.5 & 10.0 & 48.96 \\
 41,000 & 4.5 & 10.2 & 49.05 \\
 42,000 & 4.5 & 10.4 & 49.14 \\
 43,000 & 4.5 & 10.6 & 49.23 \\
 44,000 & 4.5 & 10.8 & 49.31 \\
 45,000 & 4.5 & 11.0 & 49.38 \\
\hline
\end{tabular}
\label{LTE_models}
\end{table}

\end{document}